\documentstyle[preprint,prd,aps]{revtex}
\newcommand{\be}{\begin{equation}}
\newcommand{\ee}{\end{equation}}
\newcommand{\bea}{\begin{eqnarray}}
\newcommand{\eea}{\end{eqnarray}}

\newcommand{\sptwo}{1.4}
\newcommand{\doublespace}{\edef\baselinestretch{\sptwo}\Large\normalsize}

\newcommand{\newsection}[1]{
\section{#1}
\setcounter{equation}{0}}

\renewcommand{\theequation}{\thesection.\arabic{equation}}

\newcounter{newapp}
\renewcommand{\thenewapp}{\Alph{newapp}}
\begin{document}
\draft
\title{The supercharge and superconformal symmetry for $N=1$ supersymmetric 
quantum mechanics}
\author{T.E. Clark\footnote{e-mail address: clark@physics.purdue.edu}, S.T. 
Love\footnote{e-mail address: love@physics.purdue.edu} and S.R. 
Nowling\footnote{e-mail address: nowling@uiuc.edu}}
\address{\it Department of Physics, 
Purdue University,
West Lafayette, IN 47907-1396}
\maketitle
\begin{abstract}
The superspace Lagrangian formulation of $N=1$ supersymmetric quantum mechanics 
is presented.  The general Lagrangian constructed out of chiral and antichiral 
supercoordinates containing up to two derivatives and with a canonically 
normalized  kinetic energy term  describes the motion of a nonrelativistic spin 
$1/2$ particle with Land\'e $g$-factor 2 moving in two spatial dimensions under 
the influence of a static but spatially dependent magnetic field. Noether's 
theorem is derived for the general case and is used to construct superspace 
dependent charges whose lowest components give the superconformal generators.  
The supercoordinate of charges containing an $R$ symmetry charge, the 
supersymmetry charges and the Hamiltonian are combined to form a supercharge 
supercoordinate.  Superconformal Ward identities for the quantum effective 
action are derived from the conservation equations and the source of potential 
symmetry breaking terms are identified.
\end{abstract}

\newpage
\doublespace

\newsection{Introduction}

Supersymmetric quantum mechanics \cite{Witten} has proven to be a fruitful 
testing ground for many of the intriguing consequences of the supersymmetry 
(SUSY) algebra which are also anticipated to be applicable in supersymmetric 
quantum field theories. Conversely, for those quantum mechanical models 
exhibiting a supersymmetry, the symmetry algebra can be exploited to extract 
many of the model  spectral properties \cite{Khare}. For example, the zero 
energy ground state wavefunctions for the 1 space dimension quantum mechanical 
systems whose dynamics is governed by the potential energy functions
\be
V(x)=\frac{1}{2}W^2(x)\pm\frac{1}{2}\frac{dW(x)}{dx} ~,
\ee
where $W(x)$ is an arbitary real function (superpotential), are given by 
\be
\psi_{\pm 0}(x)=Ne^{\mp\int_{x_0}^x dx^\prime W(x^\prime)} ~,
\ee
where $N$ is a normalization constant. This result can be derived using the 
underlying $N=1$ supersymmetry exhibited by models of this type. The fact that 
models with such potential energy functions possess an underlying supersymmetry 
can be demonstrated in a variety of ways. One approach which manifestly 
incorporates the $N=1$ supersymmetry from the outset, entails the construction 
of a superspace consisting of the ordinary time coordinate, an anticommuting 
Grassmann variable and its conjugate. (See Appendix A for a detailed 
construction of such a superspace and its properties). The above class of models 
then emerges from the supersymmetric invariant action 
\be
\Gamma_0=\int dV [\frac{1}{2}D\Phi \bar{D}\Phi -f(\Phi)]
\ee
where $\Phi = \Phi^\dagger$ is a real supercoordinate whose $\theta\bar\theta$ 
component is $W(x)$ and after application of the equations of motion 
$\frac{df(x)}{dx}=-W(x)$. The Hamiltonians for the two partner potential energy 
functions take the factorized form 
\bea
H_\pm &=&-\frac{1}{2}\frac{d^2}{dx^2}+\frac{1}{2}W(x)^2 \mp 
\frac{1}{2}\frac{dW(x)}{dx}\cr
&=& \frac{1}{\sqrt{2}}\left(\pm \frac{d}{dx}-
W(x)\right)\frac{1}{\sqrt{2}}\left(\mp \frac{d}{dx}-W(x)\right).
\eea 
The zero energy ground states satisfy $\frac{1}{\sqrt{2}}(\pm \frac{d}{dx}-
W(x))\psi_{\pm 0} (x)=0$ whose solution is given above.

Besides using real supercoordinates, the $N=1$ supersymmetry can also be 
realized in terms of chiral and antichiral supercoordinates. The purpose of this 
paper is to study the consequences of such realizations. In section 2, we 
present the most general supersymmetric action constructed out of such 
coordinates containing up to two derivatives. For a canonically normalized 
kinetic term, the model  reduces to that describing a nonrelativistic spin 1/2 
particle with Land\'e $g$-factor 2 moving in 2 spatial dimensions under the 
influence of a static, but spatially dependent magnetic field.  

The remainder of the paper is devoted to studying the space-time symmetry 
structure of the general $N=1$ supersymmetric action constructed out of the 
chiral and antichiral supercoordinates. The possible symmetries of the model are 
determined and the associated charges are constructed.  The symmetry 
transformations are applied to the Green's functions and their 
(non-)conservation laws are expressed as Ward identities for the quantum 
effective action.   The time translation and SUSY transformations of the 
supercoordinates are represented as superspace differential operators.  The 
superspace action is shown to be invariant under these variations.  The $N=1$ 
supersymmetry algebra and the construction of its representation on 
supercoordinates is presented in detail in Appendix A.  There the methodology 
for constructing SUSY invariant actions is prescribed.  The derivation of the 
superspace Noether's theorem for the construction of charges generating 
variations of the coordinates is presented in Section 3.  The most general 
space-time symmetries of the model are found to be those of the $OSp(2,1)$ 
superconformal group along with an additional $U(1)$ internal symmetry phase 
transformation.  The Noether charge for each symmetry is constructed and the 
conservation laws derived.  In general, the $U(1)$ and (super-)conformal 
symmetries are not conserved and the symmetry violating variation of the 
Lagrangian is presented.  The charge (non-)conservation laws are used to derive 
the Ward identities obeyed by the generating functional for one-particle 
irreducible functions.  The superconformal algebra is presented in Appendix B 
along with the representation of the symmetry transformations as superspace 
differential operator variations of the supercoordinates.  The charges are shown 
to form the components of superspace multiplets which transform into each other 
under a restricted set of SUSY transformations.  The multiplets are known as 
quasi-supercoordinates.  Appendix C recalls the quantum action principle in 
superspace that underlies the derivation of the various Ward identities for the 
effective action, while in Appendix D, the spinning coordinate path integral is 
explicitly performed by evaluating the Grassmann determinant. So doing, the 
model is recast in terms of a matrix Hamiltonian which depends only on the 
particle ordinary spatial coordinates. Finally in Section 4, the multiplet that 
contains as its components a $R$-symmetry charge, the SUSY charges and the 
Hamiltonian is shown to form a supercoordinate referred to as the supercharge.  
All of the superconformal charges are constructed from these charges along with 
terms that are related to the global Weyl scaling charge for the chiral and 
antichiral supercoordinates.  These are just the chiral, antichiral and total 
number operators.   The Ward identity operators for the superconformal and 
$U(1)$ symmetries also form (quasi-)supercoordinates. The symmetry violating 
terms are determined for all the symmetry transformations.  Only the trivial 
theory is both conformal and conformal SUSY invariant   or simultaneously $U(1)$ 
and scale invariant. 
\newpage

\newsection{$N=1$ supersymmetry and (anti-) chiral supercoordinates}

Manifestly supersymmetric quantum mechanical models are most conveniently 
formulated directly in superspace. The construction of this superspace can be 
achieved by augmenting the ordinary time variable by the introduction of an 
additional single complex Grassmann variable $\theta$, satisfying $\theta^2=0$, 
along with its complex conjugate $\bar\theta$, also satisfying 
${\bar\theta}^2=0$. Together $(t, \theta, \bar\theta )$ corresponds to a point 
in superspace. The detailed construction of this superspace along with the 
definition of SUSY covariant derivatives as well as real, chiral and anti-chiral 
supercoordinates are presented in Appendix A. In the following, we recall some 
of the salient features. 

A chiral supercoordinate $\phi (t, \theta, \bar\theta)$ and an antichiral 
supercoordinate  $\bar\phi (t, \theta, \bar\theta)$, have the component 
expansions
\bea
\phi (t, \theta, \bar\theta) &=& e^{-i\theta\bar\theta \partial_t}\left( 
z(t)+i\sqrt{2}\theta\xi (t)\right) = z(t) + i\sqrt{2}\theta\xi (t) -
i\theta\bar\theta \dot{z}(t)\cr
\bar\phi (t, \theta, \bar\theta) &=& e^{i\theta\bar\theta \partial_t}\left( 
\bar{z}(t)-i\sqrt{2}\bar\xi (t)\bar\theta \right) = \bar{z}(t) - 
i\sqrt{2}\bar\xi (t)\bar\theta +i\theta\bar\theta \dot{\bar{z}}(t) ~.
\eea
Here the components $z(t)$ and its complex conjugate $\bar{z}(t)$ are ordinary 
coordinates, while $\xi(t),~\bar\xi (t)$ are Grassmann valued (spinning) 
coordinates. (Throughout the paper, a time derivative will be denoted by 
$\dot{z}(t) = \partial_t z(t) =\frac{d}{d t}z(t)$).

The most general SUSY invariant action through 2 derivatives may then be written 
in terms of the (anti-) chiral supercoordinates as
\be
\Gamma_0 = \int dV {\cal L} = \int dV \left[ K(\phi, \bar\phi) + g(\phi, 
\bar\phi) D\phi \bar{D}\bar\phi  \right]
\label{classicalaction}
\ee
where the real K\"ahler prepotential supercoordinate $K=K(\phi, \bar{\phi})$  
and real supercoordinate $g(\phi, \bar{\phi})$ are, in general, independent 
functions of the chiral and anti-chiral supercoordinates, $\phi$ and 
$\bar{\phi}$,  but not there SUSY covariant derivatives. The nilpotent SUSY 
covariant derivatives are defined as $D=\frac{\partial}{\partial \theta}-
i\bar\theta\frac{\partial}{\partial t}$ ($D^2=0$) and $\bar{D}=-
\frac{\partial}{\partial \bar\theta}+i\theta\frac{\partial}{\partial t}$ 
($\bar{D}^2=0$). They anticommute to give the (SUSY covariant) time derivative 
$\{D, \bar{D}\}=2i\frac{\partial}{\partial t}$.
Performing the component decomposition yields the action
\bea
 \Gamma_0 &=& \int dt L = \int dt \left\{ 4 
g(z,\bar{z})\dot{z}\dot{\bar{z}}+i\left( K_z (z,\bar{z}) \dot{z} -K_{\bar 
z}(z,\bar{z})\dot{\bar{z}}\right)  \right.\cr
 & & \left.\qquad\qquad\qquad + 2ig(z,\bar{z})\left(\xi\dot{\bar\xi}-\dot\xi 
\bar\xi \right) -K_{z\bar{z}} (z,\bar{z}) [\xi,\bar\xi] \right.\cr
 & & \left.\qquad\qquad\qquad\qquad\qquad -i\left( g_z(z,\bar{z})\dot{z} -
g_{\bar z}(z,\bar{z}) \dot{\bar{z}} \right) [\xi,\bar\xi ] \right\} ~,
\label{classicalactioncomp}
\eea
Throughout the paper, subscripts indicate differentiation with respect to that 
variable so that, for example, $K_z= \frac{\partial K}{\partial z}$ , 
$K_{z\bar{z}}=\frac{\partial^2 K}{\partial z \partial \bar{z}} $. 

We first examine the above action when the kinetic energy term is canonically 
normalized which corresponds to taking $g(z, \bar{z})=\frac{1}{4}$ so the 
Lagrangian takes the form   
\be
L = \dot{z}\dot{\bar{z}} +\frac{i}{2}\left( \xi\dot{\bar\xi} - \dot\xi 
\bar\xi\right) +i\left( K_z\dot{z} -K_{\bar{z}}\dot{\bar{z}}\right) -
K_{z\bar{z}} [\xi, \bar\xi ] ~.
\label{action1}
\ee
Under the supersymmetry transformations of the coordinates given by
\bea
i[Q , z] =\delta^Q z &=& i\sqrt{2}\xi \qquad\qquad\qquad i\{Q , \xi \}=\delta^Q 
\xi = 0 \cr
i[Q , \bar{z} ]=\delta^Q \bar{z} &=& 0 \qquad\qquad\qquad i\{Q ,\bar\xi 
\}=\delta^Q \bar\xi = -\sqrt{2}\dot{\bar{z}}
\eea
\bea
i[\bar{Q} ,z]=\delta^{\bar{Q}} z &=& 0 \qquad\qquad\qquad i\{\bar{Q} , \xi 
\}=\delta^{\bar{Q}} \xi = \sqrt{2}\dot{z} \cr
i[\bar{Q} , \bar{z} ]=\delta^{\bar{Q}} \bar{z} &=& -i\sqrt{2}\bar\xi 
\qquad\qquad\qquad i\{ \bar{Q} , \bar\xi \}=\delta^{\bar{Q}} \bar\xi = 0 ~,
\eea
the Lagrangian transforms as a total time derivative
\bea
\delta^Q L &=& \frac{d}{dt}\left(\frac{i}{\sqrt{2}}\dot{\bar{z}}\xi -\sqrt{2} 
K_z \xi \right)\cr
\delta^{\bar{Q}} L &=& \frac{d}{dt}\left(-\frac{i}{\sqrt{2}}\dot{z}\bar\xi -
\sqrt{2} K_{\bar{z}} \bar\xi \right) ~.
\eea
Consequently, the action, $\Gamma_0 = \int dt L$, is SUSY invariant: $\delta^Q 
\Gamma_0 =0= \delta^{\bar{Q}}\Gamma_0$. 

Identifying the components $z(t)=\sqrt{\frac{m}{2}}\left( x(t)+iy(t) \right )$ 
and $\bar{z}(t)=\sqrt{\frac{m}{2}}\left( x(t)-iy(t) \right)$, where $x,y$ are 
the coordinates of 2-dimensional space, and rescaling the K\"ahler prepotential 
as $K\rightarrow qK$, where $q$ is the particle electric charge, this Lagrangian 
becomes
\be
L= \frac{1}{2}m\vec{v}^2 +q\vec{A}\cdot \vec{v} +\frac{i}{2}\left(\xi 
\dot{\bar\xi}-\dot\xi \bar\xi\right) +\frac{q}{2m}B(x,y) [\xi, \bar\xi ]~,
\ee
where $\vec{v}=\hat{e}_1\dot{x}+\hat{e}_2\dot{y}$ is the particle velocity. In 
fact, this Lagrangian is recognized as one describing the planar motion of a 
nonrelativistic spin 1/2 particle with Land\'e g-factor 2 moving under the 
action of a static, but spatially dependent magnetic field orientied 
perpendicular to the plane of motion: $\vec{B}=\hat{e}_3 B(x,y)$, with 
\be
B(x,y) = \epsilon_{ij} \partial_i A_j (x,y) = - \nabla^2 K(x,y)~,
\ee
and $A_i(x,y)$ the components of the vector potential. Thus the general 
supersymmetric quantum mechanical model in 2 spatial dimensions containing up to 
2 derivatives and with canonically normalized kinetic energy term is simply that 
describing the motion of  a nonrelativistic spin 1/2 particle with $g$-factor 2 
moving under the influence of a static, but spatially dependent magnetic field. 

The more familiar form for the corresponding Hamiltonian expressed solely in 
terms of the coordinates $x, y$ and their canonical momenta can be gleaned by 
integrating out the spinning variables as is explicitly performed in Appendix D. 
Alternatively, one could employ  a specific representation using the Pauli 
matrices for the spinning coordinates: $\xi = \sigma_+$ and $\bar\xi =  \sigma_-
$ so that $\{\xi, \bar\xi\}=1$ and $[\xi, \bar\xi]=\sigma_3$. Here 
\be
\sigma_+ = \pmatrix{0&1\cr
0&0} \qquad\qquad
\sigma_- = \pmatrix{0&0\cr
1&0} \qquad\qquad
\sigma_3 = \pmatrix{1&0\cr
0&-1}
\ee
are the raising, lowering  and third component Pauli matrices respectively. In 
either case, one secures the Lagrangian
\be
L= \frac{1}{2}m\vec{v}^2 +q\vec{A}\cdot \vec{v} +\frac{q}{m}B(x,y)S ~,
\label{Lff}
\ee
where $S=\frac{1}{2}\sigma_3$ is the spin operator. The corresponding 
Hamiltonian is then simply
\be
H=\frac{\left( \vec{p} -q\vec{A}(x,y)\right)^2}{2m} - \frac{q}{m}B(x,y)S ,
\label{Ham}
\ee
 Note that the Land\'e $g$-factor is 2. As noted previously \cite{CLN}-\cite{K}, 
this Hamiltonian is supersymmetric.  

The complex, nilpotent $N=1$ supersymmetry charges, $Q$ and $\bar{Q} = 
(Q)^\dagger$, can be constructed as
\bea
Q &=& \frac{i}{\sqrt{m}} \pi \sigma_+ \cr
\bar{Q} &=& -\frac{i}{\sqrt{m}} \bar\pi \sigma_- ~,
\eea
where  
\bea
\pi &\equiv&  \left( p_x -qA_x\right) -i\left(p_y -qA_y \right) \cr
\bar\pi &\equiv&  \left( p_x -qA_x\right) +i\left(p_y -qA_y \right) ,
\eea
have commutator
\be
\left[ \bar\pi, \pi \right] = 2q B(x,y) .
\ee
These $Q, \bar{Q}$ nilpotent charges are seen to obey the supersymmetry algebra 
\bea
\left\{ Q, Q\right\} = &0&=\left\{ \bar{Q}, \bar{Q}\right\}\cr
\left\{ Q, \bar{Q}\right\} &=& 2H \cr
\left[ Q, H \right] = &0& = \left[\bar{Q}, H\right] .
\label{SUSYalg}
\eea
The Hamiltonians for the two different spin projections form partners of this 
supersymmetric quantum mechanical system. The supersymmetry algebra was 
exploited \cite{CLN} to explicitly construct the exact zero energy ground state 
wavefunction for the system. Modulo this ground state, the remainder of the 
eigenstates and eigenvalues of the two partner Hamiltonians form positive energy 
degenerate pairs \cite{CLN}.

For any supercoordinate $\phi (t, \theta,\bar\theta) $, the supersymmetry 
transformations can be represented by superspace Grassmann derivatives (see 
Appendix A) as
 \bea
\delta^Q \phi (t, \theta,\bar\theta)  &=& \left(\frac{\partial}{\partial 
\theta}+i\bar\theta\frac{\partial}{\partial t}\right) \phi (t, 
\theta,\bar\theta) \cr
\delta^{\bar{Q}}\phi (t, \theta,\bar\theta)  &=& \left( -
\frac{\partial}{\partial \bar\theta}-i\theta\frac{\partial}{\partial t}\right) 
\phi (t, \theta,\bar\theta)  ~.
\eea
The SUSY variations anticommute to yield a time derivative: $\{\delta^Q, 
\delta^{\bar{Q}}\} = -2i\partial_t$.  Since time translations are generated by 
time derivatives, this gives a representation of the SUSY algebra 
(\ref{SUSYalg})  
\bea
\{\delta^Q, \delta^Q \} = &0& = \{\delta^{\bar{Q}}, \delta^{\bar{Q}}\}\cr
\{\delta^Q ,\delta^{\bar{Q}}\} &=& -2i\delta^H \cr
[\delta^Q, \delta^H] = &0&= [\delta^{\bar{Q}}, \delta^H] ,
\eea
where $\delta^H \phi (t, \theta,\bar\theta) = \partial_t \phi (t, 
\theta,\bar\theta) $. 

In addition to time translation and SUSY invariance, the action of equation 
(\ref{classicalactioncomp}) also exhibits a symmetry under a $U(1)$ phase 
transformation of the (spinning) Grassmann coordinates given by  $\xi^{~\prime}= 
e^{i\rho}\xi $ and $\bar\xi^{~\prime}=e^{-i\rho}\bar\xi$, where $\rho$ is a real 
parameter, while the coordinates $z$ and $\bar{z}$ are left invariant. Moreover, 
if $g$ and $K$ are functions only of the combination  $\bar\phi \phi$, then the 
action is also invariant under rotations in the $x-y$ plane. This $U(1)$ phase 
transformation takes the form $\phi \rightarrow \phi^{~\prime}= e^{i\alpha}\phi$ 
and $\bar\phi \rightarrow \bar\phi^{~\prime} =e^{-i\alpha}\bar\phi$. Note that, 
in this case, it is the entire multiplet, including the ordinary and Grassmann 
odd coordinates, which participate in the rotation. These transformations are 
generated by superspace differential operators. Denoting the generator of the 
$U(1)$ transformation involving the entire supercoordinate by $J$ and the 
spinning coordinate only rotation by $R_0$,  one finds 
\bea
\delta^J \phi (t, \theta,\bar\theta) &=& i\phi(t,\theta,\bar\theta) \qquad , 
\qquad 
\delta^J \bar\phi (t, \theta,\bar\theta) = -i\bar\phi(t,\theta,\bar\theta)\cr
\delta^{R_0} \phi (t, \theta,\bar\theta) &=& 
i\left(\theta\frac{\partial}{\partial\theta} -\bar\theta 
\frac{\partial}{\partial\bar\theta}\right) \phi(t,\theta,\bar\theta) \quad , 
\quad
\delta^{R_0} \bar\phi (t, \theta,\bar\theta) = 
i\left(\theta\frac{\partial}{\partial\theta} -\bar\theta 
\frac{\partial}{\partial\bar\theta}\right) \bar\phi(t,\theta,\bar\theta) ~.
\eea
The Lagrangian, and hence the action, is left unchanged by $R_0$ variations, 
$\delta^{R_0} {\cal L} = 0$,
but not, in general, by $J$-transformations, except, as stated above, when $g$ 
and $K$ are functions only of $\phi\bar\phi$.

As discussed in appendices A and B, the $R$-transformations may also be defined 
to include a non-zero intrinsic $R$-weight, $n_\phi$, for the coordinates so 
that
\bea
\delta^R \phi &=& i\left( n_\phi + \theta\frac{\partial}{\partial\theta} -
\bar\theta \frac{\partial}{\partial\bar\theta}\right) \phi \cr
\delta^R \bar\phi &=& i\left( -n_\phi + \theta\frac{\partial}{\partial\theta} -
\bar\theta \frac{\partial}{\partial\bar\theta}\right) \bar\phi .
\eea
As we shall see, for a consistent embedding in the full superconformal algebra, 
it is required that $n_\phi = -2d_\phi$, where $d_\phi$ is the the scaling 
weight of $\phi$.  Note that the $R$-transformation with $n_\phi =0$ is $R_0$. 
Further note that since $\delta^R = n_\phi \delta^J +\delta^{R_0}$, in order for 
the action to be invariant under the general $R$-transformations, it must be 
$J$-invariant.

Throughout the paper, the transformations will be treated as functional 
differential operators or as derivatives acting in superspace.  For a 
transformation ${\cal Q}$ with superspace differential operator representations 
$\delta^{\cal Q}\phi $ and $\delta^{\cal Q} \bar\phi$, the corresponding 
functional differential operator is defined as
\be
\delta^{\cal Q} \equiv \int dS \delta^{\cal Q} \phi \frac{\delta}{\delta \phi} + 
\int d \bar{S}
\delta^{\cal Q} \bar\phi \frac{\delta}{\delta \bar\phi} .
\ee
So defined, the $\delta^{\cal Q}$ variation obeys the same algebra as the 
representation of the variation of the individual coordinates $\delta^{\cal 
Q}\phi$.  The quantum action principle then relates the symmetry variations of 
the quantum effective action, $\Gamma$, to the transformations of the classical 
action $\Gamma_0$ (see Appendix C) as
\be
\delta^{\cal Q} \Gamma = \left[ i\delta^{\cal Q} \Gamma_0 \right]\Gamma = \left[ 
i\int dV \delta^{\cal Q} {\cal L} \right] \Gamma ,
\ee
where the right hand side is to be interpreted in terms of inserted one-particle 
irreducible (1-PI) functions.  Thus, if the classical action is invariant so 
that $\delta^{\cal Q} \Gamma_0 = \int dV \delta^{\cal Q}{\cal L} = 0$, then so 
is the quantum effective action: $\delta^{\cal Q} \Gamma  = 0$.  This is the 
${\cal Q}$ symmetry Ward identitly for the generating functional of one-particle 
irreducible functions. (It should be noted that for certain highly singular 
potentials, a renormalization of the Green's functions may be required 
\cite{Jackiw}. Further note that this procedure could potentially violate 
certain of the naive invariances of the classical action, a well known, yet 
anomalous, situation in field theory). 

Since the classical action is SUSY and time translation invariant so is the 
quantum effective action
\bea
\delta^Q \Gamma &=& 0 = \delta^{\bar{Q}} \Gamma \cr
\delta^H \Gamma &=& 0 .
\eea
Moreover, since the $R$-variations with zero $R$-weight are just total $\theta$ 
and $\bar\theta$ derivatives,
$\delta^{R_0} \phi = i(\theta \frac{\partial}{\partial \theta} - \bar\theta 
\frac{\partial}{\partial \bar\theta})\phi $, the action is $R_0$-invariant.  
Hence
\be
\delta^{R_0} \Gamma = \left[ i\int dV \delta^{R_0} {\cal L} \right] \Gamma = 0 ,
\ee
for $n_\phi = 0 = n_{\bar\phi}$.  On the other hand, if $K$ and $g$ are not  
functions of $\bar\phi \phi$ only, then the action is not $J$ (or $R$) invariant 
but instead takes the form
\be
\delta^J \Gamma = [i\int dV \delta^J {\cal L} ]\Gamma = [\int dV \left( 
\left(\bar\phi K_{\bar\phi} - \phi K_\phi \right) + \left(\bar\phi g_{\bar\phi} 
-\phi g_\phi \right) D\phi \bar{D}\bar\phi \right) ]\Gamma .
\ee
\newpage

\newsection{Noether's Theorem And Superconformal Transformations}

A transformation of the coordinates corresponds to a good symmetry if the action 
remains invariant.  In the present case, the action, $\Gamma_0$, is given as the 
integral over the real measure of the superspace Lagrangian as
$\Gamma_0 = \int dV {\cal L}$.  Hence, the variation of the action is zero, 
$\delta \Gamma_0 =
\int dV \delta {\cal L} = 0$, provided the Lagrangian transforms into the sum of 
chiral, $F$, and  antichiral, $\bar{F}$, terms:
$\delta {\cal L} = F +\bar{F}$. Note that since $\{D, \bar{D}\}=2i\partial_t $ 
and $D^2 =0= \bar{D}^2$, the case that the Lagrangian transforms into a total 
time derivative is subsummed by the above condition.  Noether's theorem is 
obtained by considering the variation of the Lagrangian under the 
transformations of the supercoordinates, $\phi \rightarrow \phi +\delta\phi$ and 
$\bar\phi \rightarrow \bar\phi +\delta\bar\phi $, so that the Lagrangian 
transforms as
\be
\delta L = -\delta \phi \frac{1}{\bar{D}}\frac{\delta \Gamma_0}{\delta \phi} -
\delta\bar\phi \frac{1}{D}\frac{\delta\Gamma_0}{\delta\bar\phi}   +D\left( 
\delta\phi \frac{\partial {\cal L}}{\partial D\phi}\right) +\bar{D}\left( 
\delta\bar\phi \frac{\partial {\cal L}}{\partial \bar{D}\bar\phi}\right) ,
\label{deltaL}
\ee
where the Euler-Lagrange derivatives are given by
\bea
\frac{\delta \Gamma_0}{\delta \phi} &=& -\bar{D} \left( \frac{\partial {\cal 
L}}{\partial \phi} -D \left(\frac{\partial {\cal L}}{\partial D\phi} \right) 
\right) \cr
\frac{\delta \Gamma_0}{\delta \bar\phi} &=& -{D} \left( \frac{\partial {\cal 
L}}{\partial \bar\phi} -\bar{D} \left(\frac{\partial {\cal L}}{\partial 
\bar{D}\bar\phi} \right) \right)  .
\label{EL}
\eea
Here the $1/D$ or $1/\bar{D}$ operators in (\ref{deltaL}) are understood to 
cancel the corresponding $D$ or $\bar{D}$ derivatives in (\ref{EL}).
Acting on equation (\ref{deltaL}) with $D$ and $\bar{D}$ yields
\bea
D\delta {\cal L} &=& -\delta \bar\phi \frac{\delta \Gamma_0}{\delta \bar\phi} -
\left( D\delta \bar\phi\right) 
\left(\frac{1}{D}\frac{\delta\Gamma_0}{\delta\bar\phi}\right) +D\left( \bar{D} 
\left(\delta\bar\phi \frac{\partial {\cal L}}{\partial \bar{D} \bar\phi}\right) 
-\delta\phi \left(\frac{1}{\bar{D}}\frac{\delta\Gamma_0}{\delta \phi}\right) 
\right) \cr
\bar{D}\delta {\cal L} &=& -\delta \phi \frac{\delta \Gamma_0}{\delta \phi} -
\left( \bar{D}\delta \phi\right) 
\left(\frac{1}{\bar{D}}\frac{\delta\Gamma_0}{\delta \phi}\right) +\bar{D}\left( 
{D} \left(\delta \phi \frac{\partial {\cal L}}{\partial {D} \phi}\right) -
\delta\bar\phi \left(\frac{1}{{D}}\frac{\delta\Gamma_0}{\delta \bar\phi}\right) 
\right)  .
\eea
These identities lead to the superspace Noether's theorem. The Noether charge, 
${\cal Q}$, corresponding to this transformation is defined as
\be
{\cal Q}  \equiv  \frac{1}{2} \left\{ \bar{D} \left( \delta \bar\phi 
\frac{\partial {\cal L}}{\partial \bar{D}\bar\phi}\right)  -\delta \phi \left( 
\frac{1}{\bar{D}} \frac{\delta\Gamma_0}{\delta \phi}\right)  - D\left( 
\delta\phi \frac{\partial {\cal L}}{\partial D \phi}\right)+ \delta \bar\phi 
\left( \frac{1}{D} \frac{\delta \Gamma_0}{\delta \bar\phi}\right) \right\} .
\label{NC}
\ee
It follows that the SUSY covariant derivatives of the charge are
\bea
D {\cal Q} &=& \delta \bar\phi \frac{\delta \Gamma_0}{\delta \bar\phi} + 
\frac{1}{2} D \delta {\cal L} + \left( D \delta \bar\phi\right) 
\left(\frac{1}{D} \frac{\delta \Gamma_0}{\delta \bar\phi} \right) \cr
\bar{D} {\cal Q} &=& -\delta \phi \frac{\delta \Gamma_0}{\delta \phi} - 
\frac{1}{2} \bar{D} \delta {\cal L} - \left( \bar{D} \delta \phi \right) 
\left(\frac{1}{\bar{D}} \frac{\delta \Gamma_0}{\delta \phi} \right) .
\eea
Further differentiation leads to the local form of Noether's theorem
\bea
\left\{ D, \bar{D} \right\} {\cal Q} &=& 2i \frac{\partial}{\partial t} {\cal Q} 
\cr
 &=& \bar{D} \left( \delta \bar\phi \frac{\delta \Gamma_0}{\delta \bar\phi}  + 
\left(D\delta\bar\phi\right) \frac{1}{D} \frac{\delta \Gamma_0}{\delta \bar\phi} 
\right) \cr
 & &\qquad -D\left( \delta \phi \frac{\delta \Gamma_0}{\delta \phi} + \left( 
\bar{D} \delta\phi \right) \frac{1}{\bar{D}} \frac{\delta \Gamma_0}{\delta 
\phi}\right) 
 -\frac{1}{2} \left[ D, \bar{D} \right] \delta {\cal L}  .
\eea
Inserting this in the one-particle irreducible functions, applying the equations 
of motion for the Green's functions (see Appendix C) and integrating over the 
time while ignoring total time derivatives, the Ward identity for the variation 
of the effective action is secured in the form of the quantum action principle
\bea
\delta \Gamma &\equiv & \left( \int dS \delta \phi \frac{\delta}{\delta \phi} + 
\int d \bar{S} \delta \bar\phi \frac{\delta}{\delta\bar\phi} \right) \Gamma \cr
 &=& \left[i\int dV \delta {\cal L}\right] \Gamma  -\left[i\int dS 
\left(\bar{D}\delta\phi\right) \frac{1}{\bar{D}} \frac{\delta \Gamma_0}{\delta 
\phi}  \right]\Gamma 
- \left[i\int d\bar{S}  \left(D\delta\bar\phi\right) \frac{1}{D} \frac{\delta 
\Gamma_0}{\delta \bar\phi} \right]\Gamma ,
\eea
with $\Gamma$ the quantum effective action (the one-particle irreducible 
function generating functional). In obtaining this result, we have used that 
$\int dS \chi = \int dt d\theta \chi = \int dt D\chi $ and $\int d\bar{S} 
\bar\chi = \int dt d\bar\theta \bar\chi = -\int dt \bar{D}\bar\chi $, with 
$\chi$ and $\bar\chi$ Grassmann odd functions of the supercoordinates and their 
SUSY covariant derivatives.

We now focus on transformations which correspond to a reparametrization of 
superspace:
\bea
t &\rightarrow & t^\prime (t, \theta, \bar\theta) \cr
\theta &\rightarrow & \theta^\prime (t, \theta, \bar\theta)\cr
\bar\theta &\rightarrow & \bar\theta^\prime (t, \theta, \bar\theta)
\label{ssrepara}
\eea
Note that the variations of the superspace coordinates are general functions of 
superspace.
Under such transformations, the chiral and antichiral supercoordinates transform 
as 
\bea
\phi(t,\theta, \bar{\theta})&\rightarrow & e^{i\Lambda (t^{-1},{\theta}^{-1}, 
\bar{\theta}^{-1})}\phi(t^{-1},\theta^{-1}, \bar{\theta}^{-1}) \cr
\bar\phi(t,\theta, \bar{\theta})&\rightarrow & e^{-i\bar\Lambda (t^{-
1},{\theta}^{-1}, \bar{\theta}^{-1})}\bar\phi(t^{-1},\theta^{-1}, 
\bar{\theta}^{-1}) ~,
\eea
where $\Lambda$ and $\bar{\Lambda}$ are complex independent functions of 
superspace and the inverse transformation refers to the reparametrization given 
by equation (\ref{ssrepara}) where the primed variables are to be replaced with 
unprimed ones. For infinitesimal parameters, the reparametrization takes the 
form
\bea
t & \rightarrow & t + \delta t(t, \theta, \bar\theta) \cr
\theta & \rightarrow &  \theta + \delta \theta (t, \theta, \bar\theta) \cr
\bar\theta & \rightarrow &  \bar\theta + \delta \bar\theta (t, \theta, 
\bar\theta) ,
\eea
while the chiral and anti-chiral supercoordinates vary as
\bea
\phi(t,\theta, \bar{\theta})  &\rightarrow & \phi(t,\theta, \bar{\theta})+\delta 
\phi(t,\theta,\bar{\theta})=(1+i\delta \Lambda )\phi(t-\delta t,\theta -\delta 
\theta, \bar{\theta}-\delta \bar{\theta}) \cr
\bar{\phi}(t,\theta, \bar{\theta}) & \rightarrow & \bar{\phi}(t,\theta, 
\bar{\theta})+\delta \bar{\phi}(t,\theta, \bar{\theta})=(1+i\delta \bar{\Lambda} 
)\bar{\phi}(t-\delta t,\theta -\delta \theta, \bar{\theta}-\delta \bar{\theta})
\eea
where $\delta \Lambda$ and $\delta \bar{\Lambda}$ are infinitesimal complex 
independent functions of superspace. Retaining terms through linear in the small 
variations gives
\bea
\delta \phi &=& \left( i\delta\Lambda -\delta t \partial_t -\delta\theta 
\partial_\theta - \delta\bar\theta \partial_{\bar\theta} \right) \phi \cr
 &=& \left( i\delta\Lambda -\delta{B}\partial_t -\delta \theta D\right) \phi \cr
\delta \bar\phi &=& \left( -i\delta\bar\Lambda -\delta t \partial_t -
\delta\theta \partial_\theta - \delta\bar\theta \partial_{\bar\theta} \right) 
\bar\phi \cr
 &=& \left( -i\delta\bar\Lambda -\delta{B}\partial_t + \delta\bar\theta \bar{D} 
\right) \bar\phi ,
\eea
where $\delta{B} \equiv \delta t +i\delta\theta\bar\theta +i\delta\bar\theta 
\theta  $. 

Under this reparametrization, the terms in the superspace Lagrangian, ${\cal 
L}=K +gD\phi\bar{D}\bar\phi$, transform as  
\bea
\delta(gD\phi\bar{D}\bar\phi ) &=& ig\phi\bar\phi \left(\bar{D}D\delta\Lambda + 
D\bar{D}\delta\bar\Lambda \right) \cr
 & &+i\left(\delta\Lambda -\delta\bar\Lambda +i \left(D\delta\theta\right)-i 
\left(\bar{D}\delta\bar\phi\right) \right)gD\phi\bar{D}\bar\phi \cr
 & &-ig\bar{D}\left(D\delta\Lambda \phi\bar\phi \right) -
igD\left(\bar{D}\delta\bar\Lambda \phi\bar\phi \right) +\bar{D}\left( 
\delta{B}gD\phi \dot{\bar\phi}\right) -D \left( \delta{B}g\dot\phi 
\bar{D}\bar\phi \right) \cr
 & &+\frac{i}{2} \left(\delta\Lambda - \delta\bar\Lambda\right)\left(\phi g_\phi 
+ \bar\phi g_{\bar\phi} \right) D\phi \bar{D}\bar\phi 
+\frac{i}{2}\left(\delta\Lambda + \delta\bar\Lambda\right)\left(\phi g_\phi - 
\bar\phi g_{\bar\phi} \right) D\phi \bar{D}\bar\phi 
\label{gDpDp}
\eea
\bea
\delta K &=& \frac{i}{2}\left(\delta\Lambda -\delta\bar\Lambda \right)\left(\phi 
K_\phi + \bar\phi K_{\bar\phi} \right)  +\frac{i}{2}\left(\delta\Lambda 
+\delta\bar\Lambda \right)\left(\phi K_\phi - \bar\phi K_{\bar\phi} \right) \cr
 & &+\frac{i}{2}D\left(\delta{B}\bar{D}K\right) 
+\frac{i}{2}\bar{D}\left(\delta{B} DK\right) \cr
 & &-\left( \delta\theta +\frac{i}{2}\bar{D}\delta{B}\right) DK +\left( \delta 
\bar\theta -\frac{i}{2}D\delta{B} \right) \bar{D} K ,
\eea
while the Noether charge corresponding to these variations has the form
\bea
{\cal Q} &=& \frac{1}{2} \left\{ \frac{i}{2} \left( \delta\Lambda + 
\delta\bar\Lambda \right) \left( \phi K_\phi + \bar\phi K_{\bar\phi} \right) 
+\frac{i}{2} \left( \delta\Lambda - \delta\bar\Lambda \right) \left( \phi K_\phi 
- \bar\phi K_{\bar\phi} \right)\right. \cr
 & &\left. -\delta{B}\left(\dot\phi K_\phi - \dot{\bar\phi} K_{\bar\phi} \right) 
\right. \cr
 & &\left. -\delta\theta DK -\delta\bar\theta \bar{D} K +D\left( \delta\theta 
gD\phi \bar{D}\bar\phi \right) + \delta\theta gD\phi D\bar{D}\bar\phi 
+\bar{D}\left( \delta\bar\theta gD\phi \bar{D}\bar\phi \right) -\delta\bar\theta 
g \bar{D} D\phi \bar{D}\bar\phi \right. \cr
 & &\left. -iD\left( \Lambda g \phi \bar{D}\bar\phi \right) -i \Lambda g\phi 
D\bar{D}\bar\phi +i \bar{D} \left( \bar\Lambda gD\phi \bar\phi \right) + 
i\bar\Lambda g \bar{D}D\phi \bar\phi \right. \cr
 & &\left. +D\left( \delta{B}g\dot\phi \bar{D}\bar\phi \right) 
+\delta{B}g\dot\phi D\bar{D} \bar\phi +\bar{D} \left( \delta{B} g D\phi 
\dot{\bar\phi}\right) +\delta{B} g\bar{D} D\phi \dot{\bar\phi}\right\} .
\eea

The charge generates the intrinsic variations of the coordinates as $ i[{\cal Q} 
, \phi ] = \delta^{\cal Q} \phi$. That is, it gives the change in the operator 
which results in order to insure the invariance of their matrix elements when 
compared in the two frames of reference.  From this it follows that the 
variation of a derivative of the supercoordinate is the derivative of the 
variation of the supercoordinate, e.g. $\delta^{\cal Q} D\phi = D\delta^{\cal Q} 
\phi $.  We need to insure that the symmetry charges do not change the chirality 
of the supercoordinates. This chirality consistency requirement further 
restricts the transformation parameters.  The chiral constraint, $\bar{D} \delta 
\phi = 0$, dictates that
\bea
\bar{D} \delta\Lambda &=& 0 \cr
\bar{D} \delta\theta &=& 0 \cr
\bar{D} \delta B &=& 2i\delta\theta ,
\label{chiralconstraint}
\eea
while the antichiral constraint $D\delta\bar\phi =0$ mandates 
\bea
D \delta\bar\Lambda &=& 0 \cr
D\delta\bar\theta &=& 0 \cr
D\delta{B} &=& -2i\delta \bar\theta .
\label{antichiralconstraint}
\eea
The solutions to these equations require the transformation parameters 
$\delta\Lambda,~~ \delta \theta \equiv -i\delta{X}$ 
($\delta\bar\Lambda,~~\delta\bar\theta\equiv i\delta\bar{X}$) to be  chiral 
(antichiral) functions of superspace. Note that $\delta{X}$ and $\delta\bar{X}$ 
are Grassmann odd. In addition, the equations for $\delta{B}$ are satisfied 
provided
\bea
\delta t&=&  \delta{A} + \delta{X}\bar\theta +\delta\bar{X}\theta \cr
 &=&\delta\bar{A} -\delta{X}\bar\theta -\delta\bar{X}\theta \cr
 &=& \frac{1}{2}\left(\delta{A} +\delta\bar{A}\right)
\eea
where $\delta{A} ~(\delta\bar{A}$) are chiral (antichiral): $\bar{D}\delta{A} =0 
~(D\delta\bar{A}=0)$. 

Noting that the chirality of the supercoordinate variations 
(\ref{chiralconstraint}) and (\ref{antichiralconstraint}) implies that $\delta 
\theta +\frac{i}{2}\bar{D} B =0= \delta\bar\theta -\frac{i}{2}DB$, it follows 
that the variation of the K\"ahler potential simplifies to 
\bea
\delta K &=& \frac{i}{2} \left( \delta\Lambda -
\delta\bar\Lambda\right)\left(\phi K_\phi + \bar\phi K_{\bar\phi}\right) + 
\frac{i}{2} \left( \delta\Lambda +\delta\bar\Lambda\right)\left(\phi K_\phi - 
\bar\phi K_{\bar\phi}\right) \cr
&&\quad\quad +\frac{i}{2}D\left( \delta{B}\bar{D}K\right) 
+\frac{i}{2}\bar{D}\left( \delta{B}DK\right) .
\label{dK}
\eea

Our goal is secure the form of the various parameters characterizing the 
superspace reparametrization which leads to maximum possible symmetries of the 
action without restriction to the specific forms of the $g$ and $K$ real 
superfunctions. To accomplish this, we need to further restrict the parameters 
so that the variation of the term containing $g$, Eq. ($\ref{gDpDp}$,) has an 
analogous structure to the term containing $K$, Eq. ($\ref{dK}$). This is 
achieved by requiring
\bea
\frac{\partial}{\partial t}\left( \delta\Lambda +\delta\bar\Lambda \right)=0 \cr
\delta\Lambda -\delta\bar\Lambda +D\delta{X} +\bar{D}\delta\bar{X} &=& 0 .
\label{SC}
\eea
The resultant variation of the term containing $g$ now becomes
\bea
\delta \left[ g D\phi \bar{D}\bar\phi \right] &=& \frac{i}{2}\left(\delta\Lambda 
- \delta\bar\Lambda \right)\left( \phi g_\phi + \bar\phi g_{\bar\phi} \right) 
D\phi \bar{D}\bar\phi  \cr
 & &+ \frac{i}{2} \left(\delta\Lambda + \delta\bar\Lambda \right)\left( \phi 
g_\phi - \bar\phi g_{\bar\phi} \right) D\phi \bar{D}\bar\phi \cr
 & &+\bar{D}\left( \delta{B}gD\phi \dot{\bar\phi}\right) -ig\bar{D} \left( 
D\delta\Lambda \phi \bar\phi \right) \cr
 & &-D\left( \delta{B}g\dot\phi \bar{D}\bar\phi \right) -ig D \left( 
\bar{D}\delta\bar\Lambda \phi\bar\phi \right) .
\label{dg}
\eea

Solving equations (\ref{SC}) for the transformation parameters gives
\bea
\delta\Lambda &=& \frac{1}{2}\delta\alpha +\frac{i}{2} \delta\beta +\frac{i}{2} 
\delta\gamma t +\theta \delta\lambda +\frac{1}{2}\theta\bar\theta \delta\gamma 
\cr
\delta\bar\Lambda &=& \frac{1}{2}\delta\alpha -\frac{i}{2} \delta\beta -
\frac{i}{2} \delta\gamma t +\bar\theta \delta\bar\lambda 
+\frac{1}{2}\theta\bar\theta \delta\gamma \cr
\delta{X} &=& \delta\eta +\frac{i}{2}\delta\bar\lambda t +\theta \left( 
\frac{1}{2} \delta\rho -\frac{i}{2}\delta\beta -\frac{i}{2}\delta\gamma t 
\right) +\frac{1}{2}\theta\bar\theta\delta\bar\lambda \cr
\delta\bar{X} &=& \delta\bar\eta +\frac{i}{2}\delta\lambda t +\bar\theta \left( 
\frac{1}{2} \delta\rho +\frac{i}{2}\delta\beta +\frac{i}{2}\delta\gamma t 
\right) -\frac{1}{2}\theta\bar\theta\delta\bar\lambda \cr
\delta{A} &=& \delta\epsilon -\delta\beta t -\frac{1}{2} \delta\gamma t^2 
+\theta \left( 2\delta\bar\eta +i\delta\lambda t \right) +i \theta \bar\theta 
\left( \delta\beta +\delta\gamma t \right) \cr
\delta\bar{A} &=& \delta\epsilon -\delta\beta t -\frac{1}{2} \delta\gamma t^2 -
\bar\theta \left( 2\delta\eta +i\delta\bar\lambda t \right) -i \theta \bar\theta 
\left( \delta\beta +\delta\gamma t \right) ,
\eea
where  $ \delta\alpha , \delta\beta , \delta\gamma , \delta\rho , 
\delta\epsilon$ are infinitesimal real constants and 
$\delta\eta , \delta\bar\eta , \delta\lambda , \delta\bar\lambda$ are 
infinitesimal complex constant anticommuting parameters.  Recall that 
\be
\delta{B}=\delta t +i\delta \theta \bar{\theta}+i\delta\bar\theta\theta=\frac
{1}{2}(\delta{A}+\delta\bar{A})+\delta{X}\bar\theta-\delta\bar{X}\theta ~,
\ee
while the superspace infinitesimal transformation parameters are
\bea
\delta t &=& \frac{1}{2}\left( \delta{A} + \delta\bar{A} \right) =\delta\epsilon 
-\delta\beta t -\frac{1}{2}\delta\gamma t^2 +\theta \left( \delta\bar\eta 
+\frac{i}{2} \delta\lambda t\right) -\bar\theta \left(\delta\eta 
+\frac{i}{2}\delta\bar\lambda t \right) \cr
\delta \theta &=& -i\delta{X} =-i\left(  \delta\eta 
+\frac{i}{2}\delta\bar\lambda t +\theta \left( \frac{1}{2} \delta\rho -
\frac{i}{2}\delta\beta -\frac{i}{2}\delta\gamma t \right) 
+\frac{1}{2}\theta\bar\theta\delta\bar\lambda \right) \cr
\delta \bar\theta &=& i \delta\bar{X} =i\left(  \delta\bar\eta 
+\frac{i}{2}\delta\lambda t +\bar\theta \left( \frac{1}{2} \delta\rho 
+\frac{i}{2}\delta\beta +\frac{i}{2}\delta\gamma t \right) -
\frac{1}{2}\theta\bar\theta\delta\bar\lambda  \right).
\eea
Taken together, these are recognized as the class of infinitesimal 
superconformal transformations along with an additional overall $U_J(1)$ phase 
transformation of the coordinates. To obtain the finite superconformal 
transformations, one simply integrates these equations yielding (with the 
corresponding finite parameters)
\bea
t^\prime &=& \frac{e^{-\beta} t +\epsilon +\theta\bar\eta -\bar\theta 
\eta}{1+\frac{1}{2}\gamma t -\frac{i}{2}\theta \lambda +\frac{i}{2} \bar\theta 
\bar\lambda } \cr
\theta^\prime &=& \frac{e^{-\frac{i}{2}\rho -\frac{1}{2}\beta } \theta -i\eta 
+\frac{1}{2}\bar\lambda t }{1+\frac{1}{2}\gamma t -\frac{i}{2}\theta \lambda 
+\frac{i}{2} \bar\theta \bar\lambda } \cr
\bar\theta^\prime &=& \frac{e^{+\frac{i}{2}\rho -\frac{1}{2}\beta } \bar\theta 
+i\bar\eta -\frac{1}{2}\lambda t }{1+\frac{1}{2}\gamma t -\frac{i}{2}\theta 
\lambda +\frac{i}{2} \bar\theta \bar\lambda } .
\eea

Grouping terms with common variation parameters, the infinitesimal 
superconformal and $U_J(1)$ phase transformations of the supercoordinates take 
the form
\bea
\delta \phi &=& \frac{i}{2}\delta\alpha \phi + \delta\beta\left( -\frac{1}{2} + 
t\partial_t +\frac{1}{2}\theta D \right)\phi +\frac{i}{2}\delta\rho \left( 
\theta D + 2i\theta\bar\theta \partial_t \right) \phi \cr
 & &+\frac{1}{2}\delta\gamma \left( -t +i\theta\bar\theta +t^2 \partial_t + 
t\theta D \right)\phi -\delta\epsilon \partial_t \phi \cr
 & &+i\delta\eta \left( D +2i\bar\theta\partial_t \right)\phi +\delta\bar\eta 
2\theta\partial_t \phi \cr
 & &-i\delta\lambda \theta\left(1-t\partial_t \right)\phi -\frac{1}{2} 
\delta\bar\lambda \left( tD -i\theta\bar\theta D +2i\bar\theta t\partial_t 
\right)\phi \cr
\delta \bar\phi &=&-\frac{i}{2}\delta\alpha \bar\phi + \delta\beta\left( -
\frac{1}{2} + t\partial_t -\frac{1}{2}\bar\theta \bar{D} \right)\bar\phi 
+\frac{i}{2}\delta\rho \left( \bar\theta \bar{D} + 2i\theta\bar\theta \partial_t 
\right) \bar\phi \cr
 & &+\frac{1}{2}\delta\gamma \left( -t -i\theta\bar\theta +t^2 \partial_t - 
t\bar\theta \bar{D} \right)\bar\phi -\delta\epsilon \partial_t \bar\phi \cr
 & &+i\delta\bar\eta \left( \bar{D} -2i\theta\partial_t \right)\bar\phi -
\delta\eta 2\bar\theta\partial_t \bar\phi \cr
 & &+i\delta\bar\lambda \bar\theta\left(1-t\partial_t \right)\bar\phi -
\frac{1}{2} \delta\lambda \left( t\bar{D} 
+i\theta\bar\theta \bar{D} -2i\theta t\partial_t \right)\bar\phi .
\eea
Note that these are identical to the superconformal and $U_J(1)$ phase 
transformations given in Appendix B corresponding to the case where $d_\phi = 
d_{\bar\phi} = -\frac{1}{2}$ and $n_\phi = -2d_\phi = -n_{\bar\phi}=1$.  It 
follows that the largest possible symmetry group is just the direct product 
group $OSp(2,1) \otimes U_J(1)$.  

Since the supercoordinate Lagrangian ${\cal L}=K + g D\phi \bar{D}\bar\phi$ is 
the most general (through two derivatives) SUSY and time translationally 
invariant object constructed from the chiral supercoordinates, the corresponding 
action is, in general, not invariant under these more restrictive superconformal 
transformations. Mutatis mutandis, the SUSY (and time translation) Ward 
identities $\delta^Q \Gamma = 0 = \delta^{\bar{Q}} \Gamma $ are functional 
differential equations for the quantum effective action whose solution (through 
two derivatives ) determines the classical action, $\Gamma_0 = \int dV {\cal L}$ 
to be given by ${\cal L}=K + g D\phi \bar{D}\bar\phi$ with arbitrary $K$ and 
$g$. Only for specific forms of $g$ and $K$ will this larger symmetry be 
manifest. Isolating the $U_J(1)$ transformation by taking $\delta\Lambda = 
\delta\bar\Lambda =\delta\alpha / 2$, it follows from equations (\ref{dK}) and 
(\ref{dg}) that $\delta^J \Gamma = 0$ provided $K$ and $g$  satisfy  $ K_z z -
K_{\bar{z}}\bar{z} = 0 = g_z z -g_{\bar{z}}\bar{z}$. As such, $g(\phi, 
\bar\phi)$ and $K(\phi, \bar\phi)$ are functions only of $\phi\bar\phi$.  On the 
other hand, isolating the dilataion transformation by choosing the parameters as 
$\delta\Lambda =-\delta\bar\Lambda = i\delta\beta /2 $, dilatation invariance, 
$\delta^D \Gamma = 0$ results provided $K$ and $g$ are scale independent. This 
follows, using equations (\ref{dK}) and (\ref{dg}), provided $K_z z 
+K_{\bar{z}}\bar{z} = 0 = g_z z +g_{\bar{z}}\bar{z}$ which, in turn, dictates 
that $g(\phi, \bar\phi)$ and $K(\phi, \bar\phi)$ must be are functions of $\phi/ 
\bar\phi$.  When the dilatation charge is conserved, the commutation relation 
$[H, D] =iH$ implies that $e^{i\beta D} H e^{-i\beta D} = e^\beta H$ for any 
real $\beta$.  Thus if $|E>$ is an eigenstate of $H$, $H|E> =E|E>$, then $e^{-
i\beta D}|E>$ is also an energy eigenstate with energy eigenvalue $ e^\beta E$ 
for any real $\beta$.  Consequently,  the spectrum of $H$ consists of all non-
negative real numbers with the ground state having zero energy.  All excited 
states are doubly  degenerate as a consequence of the SUSY.  Finally, we note 
that requiring both $U_J(1)$ and dilatation invariance leads to the free theory 
only. 

The variation of the Lagrangian is given by the sum of equations (\ref{dK}) and 
(\ref{dg}) and can be written as 
\bea
\delta {\cal L} &=& F + \bar{F} +\frac{i}{2}\left(\delta\Lambda -
\delta\bar\Lambda \right) \left(\left(\phi K_\phi + \bar\phi K_{\bar\phi} 
\right) + \left( \phi g_\phi + \bar\phi g_{\bar\phi} \right) D\phi 
\bar{D}\bar\phi \right) \cr
 & &+\frac{i}{2}\left(\delta\Lambda +\delta\bar\Lambda \right)  \left(\left(\phi 
K_\phi - \bar\phi K_{\bar\phi} \right) + \left( \phi g_\phi - \bar\phi 
g_{\bar\phi} \right) D\phi \bar{D}\bar\phi \right) \cr
 & &+\frac{i}{2}\bar{D} \left(\delta\Lambda - \delta\bar\Lambda \right) \left( 
\phi g_\phi + \bar\phi g_{\bar\phi} \right) D\phi \bar\phi -\frac{i}{2}\bar{D} 
\left(\delta\Lambda + \delta\bar\Lambda \right) \left( \phi g_\phi - \bar\phi 
g_{\bar\phi} \right) D\phi \bar\phi \cr
 & &-\frac{i}{2}D \left( \delta\Lambda - \delta\bar\Lambda \right) \left( \phi 
g_\phi + \bar\phi g_{\bar\phi} \right) \phi \bar{D} \bar\phi +\frac{i}{2}D 
\left( \delta\Lambda + \delta\bar\Lambda \right) \left( \phi g_\phi -\bar\phi 
g_{\bar\phi} \right) \phi \bar{D} \bar\phi ,
\label{deltaL2}
\eea
where the chiral, $F$,  and antichiral, $\bar{F}$, combinations are defined by
\bea
F &=& \frac{i}{2}\bar{D} \left(\delta{B}DK\right) -i\bar{D} \left( 
D\delta\Lambda g \phi\bar\phi \right) +\bar{D} \left(\delta{B}g D\phi 
\dot{\bar\phi} \right) \cr
\bar{F} &=& \frac{i}{2}{D} \left(\delta{B}\bar{D}K\right) -i{D} \left( 
\bar{D}\delta\bar\Lambda g \phi\bar\phi \right) -{D} \left(\delta{B}g D\dot\phi 
\bar{D}\bar\phi \right) .
\label{FandFbar}
\eea
Note that the last two terms on the right hand side of equation (\ref{deltaL2}) 
can be written as  $-iD\delta\Lambda Y -i\bar{D}\delta\bar\Lambda \bar{Y}$ where 
\bea
Y &=& \bar\phi g_{\bar\phi} \phi\bar{D}\bar\phi \cr
\bar{Y} &=& \phi g_\phi D \phi \bar\phi  .
\eea
are chiral and antichiral respectively: $\bar{D} Y =0~;~D\bar{Y} =0$. For the 
special case when  $D\delta\Lambda = \delta\lambda$ and 
$\bar{D}\delta\bar\Lambda = -\delta\bar\lambda$, where $\delta\lambda$ and 
$\delta\bar\lambda$ are constant Grassmann odd superconformal transformation 
parameters, the $Y$ and $\bar{Y}$ terms can be included as $F$ and $\bar{F}$ 
terms in $\delta {\cal L}$. Since the purely chiral and antichiral terms $F + 
\bar{F}$ do not contribute to the variation of the action, they may
be absorbed into the definition of the Noether charge ${\cal Q}$ leading to an 
improved Noether charge ${\cal Q}_I$ given by
\be
{\cal Q}_I \equiv {\cal Q} +\frac{1}{2} \left( \bar{F} - F \right) ,
\ee
with ${\cal Q}$ defined in equation (\ref{NC}).  Finally, introducing the 
combination 
\be
\Delta {\cal L} \equiv \delta {\cal L} -\left( {F}+\bar{F}\right) ,
\ee
the improved Noether charge obeys the identities
\bea
D {\cal Q}_I &=& \delta \bar\phi \frac{\delta \Gamma_0}{\delta \bar\phi} 
+\frac{1}{2}D \Delta {\cal L} \cr
\bar{D} {\cal Q}_I &=& -\delta \phi \frac{\delta \Gamma_0}{\delta \phi} -
\frac{1}{2}\bar{D} \Delta {\cal L} ,
\eea
and the conservation relation
\be
-2i \frac{\partial}{\partial t} {\cal Q}_I = D\left(\delta \phi \frac{\delta 
\Gamma_0}{\delta \phi} \right) -\bar{D} \left( \delta\bar\phi \frac{\delta 
\Gamma_0}{\delta \bar\phi}\right) +\frac{1}{2} \left[ D, \bar{D}\right] \Delta 
{\cal L} .
\ee
Integrating over the time yields the quantum action principle or Ward identity 
\be
\delta \Gamma = \left[ i\int dV \delta {\cal L} \right] \Gamma .
\ee
Thus, if the classical action is invariant, $\int dV \delta {\cal L} = 0$, so is 
the effective action $\delta \Gamma =0$.

The improved superconformal charges take the form 
\bea
{\cal Q}_I &=&\frac{1}{2}\left\{ \frac{i}{2} \left( \delta\Lambda 
+\delta\bar\Lambda \right) \left( \phi K_\phi + \bar\phi K_{\bar\phi} \right) + 
\frac{i}{2} \left( \delta\Lambda -\delta\bar\Lambda\right) \left( \phi K_\phi - 
\bar\phi K_{\bar\phi} \right) \right. \cr
 & &\left. +i \delta{B} K_{\phi \bar\phi} D\phi \bar{D}\bar\phi +4i \delta{B} g 
\dot\phi \dot{\bar\phi} -i \left( \delta\Lambda + \delta\bar\Lambda +DX - 
\bar{D} \bar{X} \right) g D\phi \bar{D} \bar\phi \right. \cr
 & &\left. -\frac{i}{2} \left( \delta\Lambda +\delta\bar\Lambda \right) \left( 
\phi g_\phi + \bar\phi g_{\bar\phi} \right) D\phi \bar{D}\bar\phi -\frac{i}{2} 
\left( \delta\Lambda -\delta\bar\Lambda \right) \left( \phi g_\phi - \bar\phi 
g_{\bar\phi} \right) D\phi \bar{D}\bar\phi \right. \cr
 & &\left. +2 \left(\delta\Lambda +\delta\bar\Lambda \right) g \left( \phi 
\dot{\bar\phi} - \dot\phi \bar\phi \right) +2 \left(\delta\Lambda -
\delta\bar\Lambda \right) g \left( \phi \dot{\bar\phi} + \dot\phi \bar\phi 
\right) \right. \cr
 & &\left. -2i D\delta\Lambda g \phi \bar{D}\bar\phi +2i 
\bar{D}\delta\bar\Lambda g D \phi \bar\phi -iD \delta\Lambda g_{\bar\phi} 
\bar\phi \phi \bar{D}\bar\phi +i \bar{D} \delta\Lambda g_\phi \phi D \phi 
\bar\phi \right. \cr
 & &\left. +2\left( \dot{\delta\bar\Lambda} -\dot{\delta\Lambda} \right) 
g\phi\bar\phi +4 \delta{X} g D\phi \dot{\bar\phi} + 4\delta\bar{X} g \dot\phi 
\bar{D} \bar\phi \right\} .
\eea
Substituting the explicit expressions for the superspace dependent 
transformation parameters and isolating the coefficients of the various c-number 
and Grassmann odd parameters, the superconformal charges (defined with the 
transformation parameters removed from their definitions) and (non-) 
conservation laws can be extracted. Below we compile the various superspace 
dependent (improved) charges whose $\theta,~\bar\theta$ independent components 
generate the associated superconformal symmetries. In addition, we give the SUSY 
covariant derivatives of these superspace charges, their (non-) conservation 
laws and the form of the explicit breaking terms.

Throughout this section as well as in section 4, we adapt the notation that $H, 
Q, \bar{Q}, K, S, \bar{S}, R$ and $J$ represent an entire function of superspace 
whose $\theta, \bar\theta$ independent component is the actual superconformal 
and $U_J(1)$ generator. Elsewhere in the paper, these same symbols are used to 
represent just the actual generators.
 
\begin{enumerate}
\item  Time Translations: Parameter $\delta\epsilon$\\
\bea
{\cal Q}_I^H=H &\equiv & 4g \dot\phi \dot{\bar\phi} + K_{\phi\bar\phi} D\phi 
\bar{D}\bar\phi \cr
DH &=& 2i \dot{\bar\phi} \frac{\delta \Gamma_0}{\delta \bar\phi} \cr
\bar{D} H &=& -2i \dot\phi \frac{\delta \Gamma_0}{\delta \phi} \cr
\frac{\partial}{\partial t} H &=& -D\left( \dot\phi \frac{\delta 
\Gamma_0}{\delta \phi}\right) + \bar{D} \left( \dot{\bar\phi} \frac{\delta 
\Gamma_0}{\delta \bar\phi} \right) \cr
\frac{1}{2}\left[ D, \bar{D}\right] H &=& -i \left(D\left( \dot\phi \frac{\delta 
\Gamma_0}{\delta \phi}\right) + \bar{D} \left( \dot{\bar\phi} \frac{\delta 
\Gamma_0}{\delta \bar\phi} \right)\right) \cr
\Delta^H {\cal L} &=& 0 .
\label{HQ}
\eea
\item  Supersymmetry: Parameters $\delta\eta$ and $\delta\bar\eta$
\bea
{\cal Q}_I^Q &=& Q-2i\bar\theta H \cr
{\cal Q}_I^{\bar{Q}} & = & \bar{Q} +2i \theta H \cr
\frac{\partial}{\partial t} {\cal Q}_I^Q &=& -\left( D\left( \delta^Q \phi 
\frac{\delta \Gamma_0}{\delta \phi}\right) -\bar{D} \left( \delta^Q \bar\phi 
\frac{\delta \Gamma_0}{\delta \bar\phi} \right) \right) \cr
\frac{\partial}{\partial t}{\cal Q}_I^{ \bar{Q}} &=& -\left( D\left( 
\delta^{\bar{Q}} \phi \frac{\delta \Gamma_0}{\delta \phi}\right) -\bar{D} \left( 
\delta^{\bar{Q}} \bar\phi \frac{\delta \Gamma_0}{\delta \bar\phi} \right) 
\right) \cr
\Delta^{{\cal Q}_I^Q} {\cal L} = &0& = \Delta^{{\cal Q}_I^{\bar{Q}}} {\cal L} ,
\label{SUSYQ1}
\eea
where
\bea
Q &\equiv & -4g D\phi \dot{\bar\phi} \cr
\bar{Q} &\equiv & -4g \dot\phi \bar{D} \bar\phi \cr
DQ &= & 0  =\bar{D}\bar{Q}\cr
\bar{D} Q &=& -2iH -2iD\phi \frac{\delta \Gamma_0}{\delta \phi} \cr
D \bar{Q} &=& -2iH +2i\bar{D} \bar\phi \frac{\delta \Gamma_0}{\delta \bar\phi} 
\cr
\frac{\partial}{\partial t} Q &=& -\left\{ D\left( \delta^Q \phi \frac{\delta 
\Gamma_0}{\delta \phi}\right) -\bar{D} \left( \delta^Q \bar\phi \frac{\delta 
\Gamma_0}{\delta \bar\phi} \right) \right\} \cr
 & & -2i\bar\theta \left\{ D\left(\dot\phi \frac{\delta \Gamma_0}{\delta \phi} 
\right) -\bar{D} \left( \dot{\bar\phi} \frac{\delta \Gamma_0}{\delta 
\bar\phi}\right)\right\} \cr
\frac{\partial}{\partial t} \bar{Q} &=& -\left\{ D\left( \delta^{\bar{Q}} \phi 
\frac{\delta \Gamma_0}{\delta \phi}\right] -\bar{D} \left[ \delta^{\bar{Q}} 
\bar\phi \frac{\delta \Gamma_0}{\delta \bar\phi} \right) \right\} \cr
 & & +2i\theta \left\{ D\left(\dot\phi \frac{\delta \Gamma_0}{\delta \phi} 
\right) -\bar{D} \left( \dot{\bar\phi} \frac{\delta \Gamma_0}{\delta 
\bar\phi}\right)\right\} .
\label{SUSYQ2}
\eea
(Note that a in slight variation from the usual notational convention $Q$ is 
antichiral, $(DQ=0)$, while $\bar{Q}$ is chiral, $(\bar{D}\bar{Q} =0)$.)
\item  $U_J(1)$ Transformations: Parameter $\delta\alpha$
\bea
{\cal Q}_I^{U(1)} &=& J \equiv -2i g D\phi \bar{D}\bar\phi +4g \left(\phi 
\dot{\bar\phi} -\dot\phi \bar\phi \right) -i \left( \phi g_\phi + \bar\phi 
g_{\bar\phi} \right) D\phi \bar{D} \bar\phi +i \left( \phi K_\phi + \bar\phi 
K_{\bar\phi} \right) \cr
DJ &=& -2i \bar\phi \frac{\delta \Gamma_0}{\delta \bar\phi} + \frac{1}{2}D 
\Delta^{U(1)}{\cal L}  \cr
\bar{D} J &=& -2i\phi \frac{\delta \Gamma_0}{\delta \phi} -\frac{1}{2} \bar{D} 
\Delta^{U(1)}{\cal L}  \cr
\frac{\partial}{\partial t} J &=& -D\left( \phi \frac{\delta \Gamma_0}{\delta 
\phi}\right) -\bar{D} \left( \bar\phi \frac{\delta \Gamma_0}{\delta 
\bar\phi}\right) +\frac{i}{4} \left[ D , \bar{D} \right] \Delta^{U(1)}{\cal L} 
\cr
\Delta^{U(1)} {\cal L} &=& 2i \left\{ \left( \phi K_\phi - \bar\phi 
K_{\bar\phi}\right) + \left( \phi g_\phi - \bar\phi g_{\bar\phi} \right) D\phi 
\bar{D}\bar\phi \right\} .
\label{U(1)Q}
\eea
\item  $R$ Transformations: Parameter $n_\phi \delta\alpha + \delta\rho$
\bea
{\cal Q}_I^R &=& \frac{1}{2}\left(R -\theta Q -\bar\theta \bar{Q} +2i 
\theta\bar\theta H \right) \cr
2i\frac{\partial}{\partial t} {\cal Q}_I^R &=& -D \left( \delta^R \phi 
\frac{\delta \Gamma_0}{\delta \phi}\right) +\bar{D} \left( \delta^R \bar\phi 
\frac{\delta \Gamma_0}{\delta \bar\phi} \right) -\frac{1}{2} \left[ D , \bar{D} 
\right] \Delta^{R} {\cal L}  \cr
\Delta^{R} {\cal L} &=& \frac{n_\phi}{2} \Delta^{U(1)} {\cal L} = i n_\phi 
\left\{ \left( \phi K_\phi -\bar\phi K_{\bar\phi}\right) + \left( \phi g_\phi - 
\bar\phi g_{\bar\phi}\right)D\phi \bar{D}\bar\phi \right\} ,
\label{RQ1}
\eea
where
\bea
R & \equiv & n_\phi J + R_0 \cr
 &=& -2i (1+n_\phi ) g D\phi \bar{D}\bar\phi +4n_\phi g \left(\phi 
\dot{\bar\phi} -\dot\phi \bar\phi \right)\cr
 & &\qquad\qquad -in_\phi  \left( \phi g_\phi + \bar\phi g_{\bar\phi} \right) 
D\phi \bar{D} \bar\phi +in_\phi \left( \phi K_\phi + \bar\phi K_{\bar\phi} 
\right) 
\eea
For $n_\phi =0$, the conserved  supercoordinate charge $R_0$ is obtained as
\bea
R_0 &\equiv & -2ig D\phi \bar{D}\bar\phi \cr
DR_0 &=& Q \cr
\bar{D} R_0 &=& -\bar{Q} \cr
2i\frac{\partial}{\partial t} R_0 &=& \{ D , \bar{D} \} R_0 = -2i \left( D\phi 
\frac{\delta \Gamma_0}{\delta \phi} +\bar{D}\bar\phi \frac{\delta 
\Gamma_0}{\delta \bar\phi} \right) \cr
\frac{1}{2} \left[ D , \bar{D} \right] R_0 &=& 2i H +i \left(D\phi \frac{\delta 
\Gamma_0}{\delta \phi}  - \bar{D}\bar\phi \frac{\delta \Gamma_0}{\delta 
\bar\phi} \right)  .
\label{RQ2}
\eea

It should be pointed out that $R_0$ may also be obtained by directly applying 
Noether's theorem to the case that the supercoordinate variations are covariant 
derivatives (in which case that the transformations do not preserve the original 
chirality of the supercoordinates).  Considering the infinitesimal 
transformations 
$\delta \phi = \delta\zeta D\phi $ and $ \delta \bar\phi = \delta\bar\zeta 
\bar{D} \bar\phi$ , where $\delta\zeta$ and $\delta\bar\zeta$ are independent 
complex anticommuting parameters, equation (\ref{NC}) yields the associated 
Noether charge
\be
{\cal Q} = \left( \delta\zeta D - \delta\bar\zeta \bar{D} \right) \left( gD\phi 
\bar{D} \bar\phi + \frac{1}{2} K \right) ,
\ee
while equation (\ref{deltaL}) gives the variation of ${\cal L}$ as
\be
\delta {\cal L} = \left( \delta\zeta D + \delta\bar\zeta \bar{D} \right) K .
\ee
Absorbing $DK$ and $\bar{D}K$ into ${\cal Q}$ defines the improved charge ${\cal 
Q}_I$ as
\bea
{\cal Q}_I &=& {\cal Q} - \frac{1}{2} \left( \delta\zeta DK -\delta\bar\zeta 
\bar{D} K \right) \cr
 &=& \frac{i}{2} \left( \delta\zeta D - \delta\bar\zeta \bar{D} \right) R_0 = 
\frac{i}{2}\left( \delta\zeta Q +\delta\bar\zeta \bar{Q}\right) .
\eea
The derivative equations are easily obtained as
\bea
D {\cal Q}_I &=& \frac{i}{2}\delta\bar\zeta D\bar{D} R_0 = -
\frac{i}{2}\delta\bar\zeta D\bar{Q} = -\delta\bar\zeta \left( H- \bar{D} 
\bar\phi \frac{\delta \Gamma_0}{\delta \bar\phi} \right) \cr
\bar{D}{\cal Q}_I &=& -\frac{i}{2}\delta\zeta \bar{D}DR_0 = -
\frac{i}{2}\delta\zeta \bar{D}Q = -\delta\zeta \left( H+ {D} \phi \frac{\delta 
\Gamma_0}{\delta \phi} \right) .
\eea
leading to the time derivative 
\be
\frac{\partial}{\partial t} R_0 = - \left( D\phi \frac{\delta \Gamma_0}{\delta 
\phi} +\bar{D}\bar\phi \frac{\delta \Gamma_0}{\delta \bar\phi} \right) .
\ee

In a similar manner, the chiral and antichiral supercoordinate counting 
densities can be found by considering the number operator variations of the 
supercoordinates (global Weyl scaling)
\bea
\delta^N \phi =\phi \qquad \qquad & & \delta^{\bar{N}} \phi = 0 \cr
\delta^N \bar\phi =0 \qquad \qquad & & \delta^{\bar{N}} \bar\phi = \bar\phi .
\eea
Noether's theorem yields the counting charges 
\bea
{\cal Q}^N &=& \frac{1}{2}\left( -gD\phi \bar{D}\bar\phi -\phi g_\phi D\phi 
\bar{D}\bar\phi +\phi K_\phi -4ig \phi \dot{\bar\phi}\right)\cr
{\cal Q}^{\bar{N}} &=& \frac{1}{2}\left( +gD\phi \bar{D}\bar\phi +\bar\phi 
g_{\bar\phi} D\phi \bar{D}\bar\phi -\bar\phi K_{\bar\phi} -4ig \dot\phi 
\bar\phi\right) 
\eea
and SUSY covariant derivatives
\bea
D{\cal Q}^N &=& \frac{1}{2} D \delta^N {\cal L} \cr
\bar{D} {\cal Q}^N &=& -\phi \frac{\delta \Gamma_0}{\delta \phi} -\frac{1}{2} 
\bar{D} \delta^N {\cal L} \cr
D {\cal Q}^{\bar{N}} &=& \bar\phi \frac{\delta \Gamma_0}{\delta \bar\phi} 
+\frac{1}{2} D \delta^{\bar{N}} {\cal L} \cr
\bar{D}{\cal Q}^{\bar{N}} &=& -\frac{1}{2} \bar{D} \delta^{\bar{N}} {\cal L} ,
\eea
where
\bea
\delta^N {\cal L} &=& \frac{i}{2}R_0 +\phi K_\phi + \phi g_\phi D\phi \bar{D} 
\bar\phi \cr
 &=&\left( 1 + \phi (\ln{g})_\phi \right) gD\phi \bar{D} \bar\phi + \phi 
(\ln{K})_\phi K \cr
\delta^{\bar{N}} {\cal L} &=&\frac{i}{2}R_0 +\bar\phi K_{\bar\phi} + \bar\phi 
g_{\bar\phi} D\phi \bar{D} \bar\phi \cr
 &=& \left( 1 + \bar\phi (\ln{g})_{\bar\phi} \right) gD\phi \bar{D} \bar\phi + 
\bar\phi (\ln{K})_{\bar\phi} K .
\eea
Thus the (non-) conservation equations for the counting density are given by
\bea
\{ D, \bar{D} \} {\cal Q}^N &=& 2i \frac{\partial}{\partial t} {\cal Q}^N \cr
 &=& - D \left( \phi \frac{\delta \Gamma_0}{\delta \phi} \right) -\frac{1}{2} 
\left[ D, \bar{D} \right] \delta^N {\cal L} \cr
\{ D, \bar{D} \} {\cal Q}^{\bar{N}} &=& 2i \frac{\partial}{\partial t} {\cal 
Q}^{\bar{N}} \cr
 &=&  \bar{D} \left( \bar\phi \frac{\delta \Gamma_0}{\delta \bar\phi} \right) -
\frac{1}{2} \left[ D, \bar{D} \right] \delta^{\bar{N}} {\cal L} .
\eea
Combining these chiral and antichiral supercoordinate counting equations, one 
secures the total supercoordinate number operator (global Weyl scaling) 
conservation identity:
\bea
\frac{\partial}{\partial t} U &\equiv & 2i \frac{\partial}{\partial t} \left( 
{\cal Q}^N + {\cal Q}^{\bar{N}} \right)\cr
 &=& -D \left( \phi \frac{\delta \Gamma_0}{\delta \phi} \right) + \bar{D} \left( 
\bar\phi \frac{\delta \Gamma_0}{\delta \bar\phi} \right) +\frac{1}{2} \left[ D, 
\bar{D}\right] \left( \delta^N {\cal L} + \delta^{\bar{N}} {\cal L} \right).
\label{Udot}
\eea

This may be alternatively expressed by considering the chiral and antichiral 
counting variations of the Lagrangian directly.  Introducing the chiral  and 
antichiral combinations
\bea
T &=& -ig \phi \bar{D}\bar\phi ~~;~~(\bar{D}T=0)\cr
\bar{T} &=& -ig D\phi \bar\phi ~~;~~(D \bar{T} =0),
\eea
with the derivative identities
\bea
DT &=& -i \delta^N {\cal L} -i \frac{1}{\bar{D}} \left( \phi \frac{\delta 
\Gamma_0}{\delta \phi} \right) \cr
\bar{D} \bar{T} &=& i \delta^{\bar{N}} {\cal L} +i \frac{1}{D} \left( \bar\phi 
\frac{\delta \Gamma_0}{\delta \bar\phi} \right) ,
\eea
the total number density can be written as
\bea
U &=& 4g \frac{\partial}{\partial t} (\phi \bar\phi) + i (\phi K_\phi -\bar\phi 
K_{\bar\phi}) -i(\phi g_\phi -\bar\phi g_{\bar\phi})D\phi \bar{D}\bar\phi  \cr
 &=& DT + \bar{D}\bar{T} +i \left\{ \frac{1}{D}\left(\bar\phi \frac{\delta 
\Gamma_0}{\delta \bar\phi}\right) - \frac{1}{\bar{D}} \left( \phi \frac{\delta 
\Gamma_0}{\delta \phi} \right) \right\} .
\eea
The conservation equations take the form
\bea
DU &=& 2i \bar\phi \frac{\delta \Gamma_0}{\delta \bar\phi} + iD \delta^U {\cal 
L} \cr
\bar{D}U &=& -2i \phi \frac{\delta \Gamma_0}{\delta \phi} - i\bar{D} \delta^U 
{\cal L} ,
\eea
from which the time derivative is obtained as usual reproducing equation 
(\ref{Udot}) 
\be
-\frac{\partial}{\partial t} U = D\left( \phi \frac{\delta \Gamma_0}{\delta 
\phi} \right) -\bar{D} \left( \bar\phi \frac{\delta \Gamma_0}{\delta \bar\phi} 
\right) + \frac{1}{2} \left[ D, \bar{D} \right] \delta^U {\cal L} ,
\ee
where
\be
\delta^U {\cal L} = 2i \left( \delta^N {\cal L} + \delta^{\bar{N}} {\cal L} 
\right) .
\ee
So defined, these are the counting operators which enter into the scaling and 
conformal SUSY charges.

\item Dilatations: Parameter $\delta\beta$
\bea
{\cal Q}_I^D &=& -\frac{i}{2} \left( -tH +\frac{1}{2} U + \frac{1}{2} \theta Q -
\frac{1}{2} \bar\theta \bar{Q} \right) \cr
2i \frac{\partial}{\partial t} {\cal Q}_I^D &=& D \left( \delta^D \phi 
\frac{\delta \Gamma_0}{\delta \phi} \right) -\bar{D} \left( \delta^D \bar\phi 
\frac{\delta \Gamma_0}{\delta \bar\phi} \right) -\frac{1}{2} \left[ D, \bar{D} 
\right] \Delta^D {\cal L} \cr
\Delta^D {\cal L}  &=& -\frac{i}{4} \left( \delta^U {\cal L} +2R_0 \right)  \cr
&=& \frac{1}{2} \left( \left( \phi K_\phi + \bar\phi K_{\bar\phi} \right) + 
\left( \phi g_\phi + \bar\phi g_{\bar\phi} \right) D\phi \bar{D} \bar\phi 
\right) .
\eea
\item Superconformal Symmetry: Parameters $\delta\lambda$ and 
$\delta\bar\lambda$
\bea
{\cal Q}_I^S &=& -\frac{i}{2} t\bar{Q} +T -i \bar{D} (g\phi \bar\phi ) -\theta ( 
2i {\cal Q}_I^D +\frac{1}{2} R ) \cr
 &=& -\frac{i}{2} t\bar{Q} +2T -i Y -\theta ( 2i {\cal Q}_I^D +\frac{1}{2} R ) 
\cr
{\cal Q}_I^{\bar{S}} &=& -\frac{i}{2} tQ +\bar{T} -i D (g\phi \bar\phi ) 
+\bar\theta ( 2i {\cal Q}_I^D -\frac{1}{2} R ) \cr
 &=& -\frac{i}{2} tQ +2\bar{T} -i\bar{Y} +\bar\theta ( 2i {\cal Q}_I^D -
\frac{1}{2} R ) \cr
2i \frac{\partial}{\partial t} {\cal Q}_I^S &=& D \left( \delta^S \phi 
\frac{\delta \Gamma_0}{\delta \phi} \right) -\bar{D} \left( \delta^S \bar\phi 
\frac{\delta \Gamma_0}{\delta \bar\phi} \right)
-\frac{1}{2} \left[ D, \bar{D} \right] \Delta^S {\cal L} \cr
2i \frac{\partial}{\partial t} {\cal Q}_I^{\bar{S}}&=& D \left( \delta^{\bar{S}} 
\phi \frac{\delta \Gamma_0}{\delta \phi} \right) -\bar{D} \left( 
\delta^{\bar{S}} \bar\phi \frac{\delta \Gamma_0}{\delta \bar\phi} \right) -
\frac{1}{2} \left[ D, \bar{D} \right] \Delta^{\bar{S}} {\cal L} \cr
\Delta^S {\cal L} &=& -2i \bar\phi g_{\bar\phi} \phi \bar{D}\bar\phi -\theta ( 
2i\delta^N {\cal L} + R_0 ) \cr
 &=& -2iY -\theta ( 2i\delta^N {\cal L} + R_0 ) \cr
\Delta^{\bar{S}} {\cal L} &=& 2i \phi g_{\phi} D\phi \bar\phi +\bar\theta ( 
2i\delta^{\bar{N}} {\cal L} + R_0 ) \cr
 &=& 2i\bar{Y} +\bar\theta ( 2i\delta^{\bar{N}} {\cal L} + R_0 ) .
\eea
As alluded to in the discussion following equations (\ref{deltaL2}) and 
(\ref{FandFbar}), as a consequence of  their chiral and antichiral nature, the 
superconformal symmetry breaking terms due to $Y$ and $\bar{Y}$ can be absorbed 
into the charges to form new, improved charges.  Defining these as ${\cal 
Q}^S_{NI} = {\cal Q}_I^S +iY $ and ${\cal Q}_{NI}^{\bar{S}} ={\cal 
Q}_I^{\bar{S}} +i\bar{Y}$, the associated conservation equations are given by
\bea
2i \frac{\partial}{\partial t} {\cal Q}_{NI}^S &=& D \left( \delta^S \phi 
\frac{\delta \Gamma_0}{\delta \phi} \right) -\bar{D} \left( \delta^S \bar\phi 
\frac{\delta \Gamma_0}{\delta \bar\phi} \right)
-\frac{1}{2} \left[ D, \bar{D} \right] \delta^{S_{NI}} {\cal L} \cr
2i \frac{\partial}{\partial t} {\cal Q}_{NI}^{\bar{S}}&=& D \left( 
\delta^{\bar{S}} \phi \frac{\delta \Gamma_0}{\delta \phi} \right) -\bar{D} 
\left( \delta^{\bar{S}} \bar\phi \frac{\delta \Gamma_0}{\delta \bar\phi} \right) 
-\frac{1}{2} \left[ D, \bar{D} \right] \delta^{{\bar{S}}{NI}} {\cal L}  ,
\eea
where the breaking terms are now just
\bea
\delta^{S{NI}} {\cal L} &=& -\theta ( 2i\delta^N {\cal L} + R_0 ) \cr
\delta^{{\bar{S}}{NI}} {\cal L} &=&\bar\theta ( 2i\delta^{\bar{N}} {\cal L} + 
R_0 ) .
\eea
\item Conformal Symmetry: Parameter $\delta\gamma$
\bea
{\cal Q}_I^K &=& \frac{i}{2}\left( tU -t^2 H\right) -2ig\phi\bar\phi -\theta 
{\cal Q}_I^{\bar{S}} +\bar\theta {\cal Q}_I^S -\frac{1}{2} \theta\bar\theta R 
\cr
 &=& \frac{i}{2} \left\{ tU -t^2 H -4g \phi\bar\phi -i \theta\bar\theta R 
\right.\cr
 & &\qquad \left.+\theta \left(tQ + 4gD\phi \bar\phi +2 \phi g_\phi D\phi 
\bar\phi \right) \right. \cr
 & &\qquad\qquad \left. - \bar\theta \left( t\bar{Q} +4g \phi \bar{D} \bar\phi 
+2 \bar\phi g_{\bar\phi} \phi \bar{D} \bar\phi \right) \right\} \cr
2i \frac{\partial}{\partial t} {\cal Q}_I^K &=& -D \left( \delta^K \phi 
\frac{\delta \Gamma_0}{\delta \phi} \right) +\bar{D} \left( \delta^K \bar\phi 
\frac{\delta \Gamma_0}{\delta \bar\phi} \right) -\frac{1}{2} \left[ D, \bar{D} 
\right] \delta^K {\cal L} \cr
\delta^K {\cal L}  &=& -2t \delta^D {\cal L} -\theta \delta^{\bar{S}} {\cal L} + 
\bar\theta \delta^S {\cal L} -\frac{1}{2} \theta\bar\theta \delta^{U(1)} {\cal 
L}  \cr
 &=& -2t \delta^D {\cal L} -2i \theta \phi g_\phi D\phi \bar\phi -2i \bar\theta 
\bar\phi g_{\bar\phi} \phi \bar{D} \bar\phi  +\frac{1}{2} \theta\bar\theta 
\delta^{U(1)} {\cal L}  \cr
  &=& -2t \delta^D {\cal L} -2i \theta \bar{Y} -2i \bar\theta Y +\frac{1}{2} 
\theta\bar\theta \delta^{U(1)} {\cal L}  .
\eea
A new, improved conformal charge can also be defined by absorbing part of the 
$Y$ and $\bar{Y}$ breaking terms into the charge.  These terms in turn will 
become part of the new, improved superconformal charges replacing the improved 
ones in the conformal charge as 
\be
{\cal Q}_{NI}^K = \frac{i}{2}\left( tU -t^2 H\right) -2ig\phi\bar\phi -\theta 
{\cal Q}_{NI}^{\bar{S}} +\bar\theta {\cal Q}_{NI}^S -\frac{1}{2} 
\theta\bar\theta R .
\ee
However there still remains  $Y$ and $\bar{Y}$ dependent terms in the conformal 
symmetry (non-)conservation equation:
\bea2i \frac{\partial}{\partial t} {\cal Q}_{NI}^K &=& -D \left( \delta^K \phi 
\frac{\delta \Gamma_0}{\delta \phi} \right) +\bar{D} \left( \delta^K \bar\phi 
\frac{\delta \Gamma_0}{\delta \bar\phi} \right) -\frac{1}{2} \left[ D, \bar{D} 
\right] \Delta_{NI}^K {\cal L} -2i DY -2i \bar{D} \bar{Y}  \cr
\Delta_{NI}^K {\cal L}  &=& -2t \Delta^D {\cal L}  +\frac{1}{2} \theta\bar\theta 
\Delta^{U(1)} {\cal L}  .
\eea

\end{enumerate}

By appealing to the action principle, the (local) variations of the effective 
action, and, hence, the Green's functions, are given by the above equations.  
For general variations $\delta \phi$ and $\delta \bar\phi$, the action principle 
yields
\bea
i\left[ \delta \phi (t,\theta,\bar\theta) \frac{\delta \Gamma_0}{\delta \phi 
(t,\theta,\bar\theta)} \right] \Gamma &=& \delta \phi(t,\theta,\bar\theta) 
\frac{\delta}{\delta \phi (t,\theta,\bar\theta) } \Gamma  \cr
i\left[ \delta \bar\phi (t,\theta,\bar\theta) \frac{\delta \Gamma_0}{\delta 
\bar\phi (t,\theta,\bar\theta)} \right] \Gamma &=& \delta 
\bar\phi(t,\theta,\bar\theta) \frac{\delta}{\delta \bar\phi 
(t,\theta,\bar\theta) } \Gamma  .
\eea
Introducing the Ward identity functional differential operators for each of the 
variations of the superconformal and $U(1)$ transformations, ${\cal Q} \in \{H, 
Q, \bar{Q}, R , J, D, K, S , \bar{S} \}$,
\be
\delta^{\cal Q} \equiv \int dS \delta^{\cal Q} \phi \frac{\delta}{\delta \phi} + 
\int d \bar{S} \delta^{\cal Q} \bar\phi \frac{\delta}{\delta \bar\phi}  ,
\ee
and the local chiral and antichiral Ward identity operators
\bea
w^{\cal Q} &\equiv & D \left(\delta^{\cal Q} \phi \frac{\delta}{\delta 
\phi}\right) \cr
\bar{w}^{\cal Q} &\equiv & \bar{D} \left(\delta^{\cal Q} \bar\phi 
\frac{\delta}{\delta \bar\phi}\right) ,
\eea
Noether's theorem has the general form
\be
w^{\cal Q} \Gamma - \bar{w}^{\cal Q} \Gamma = 2 \frac{\partial}{\partial t} 
{\cal Q}_I -\frac{i}{2} \left[ D, \bar{D} \right] \Delta^{{\cal Q}} {\cal L}  .
\ee
Integrating over the time, and discarding total time differentials secures the 
quantum action principle
\be
\delta^{\cal Q} \Gamma = \left[ \int dV \Delta^{{\cal Q}} {\cal L} \right] 
\Gamma ,
\ee
where once again
\be
\delta^{\cal Q} = \int dt \left( w^{\cal Q} - \bar{w}^{\cal Q} \right) = \int dS 
\delta^{\cal Q} \phi \frac{\delta}{\delta \phi} + \int d \bar{S} \delta^{\cal Q} 
\bar\phi \frac{\delta}{\delta \bar\phi} .
\ee
The variations of the Green's functions under superconformal and $U(1)$  
transformations are obtained as Ward identity functional differential equations 
for the effective action.  Alternatively, the operator inserted Green's 
functions determine the matrix elements of the operators through the reduction 
formalism.  The local Ward identity functional differential operator terms 
vanish asymptotically, hence Noether's theorem can be viewed as an operator 
relation between the time derivative of the charge and the associated variation 
of the Lagrangian
\be
2i \frac{\partial}{\partial t} {\cal Q}_I = -\frac{1}{2} \left[ D, \bar{D} 
\right] \Delta^{{\cal Q}} {\cal L} .
\ee

\newpage

\newsection{Supercharges}

As discussed in the previous section, it is the  $\theta$ and $\bar\theta$ 
independent components of the superspace dependent Noether charges which serve 
as the superconformal and $U_J(1)$ generators. Although these Noether charges 
are functions of superspace, they do not all transform as supercoordinates. 
While some are supercoordinates, others involve the product of explicit factors 
of superspace, $(t, \theta, \bar\theta)$, multiplying supercoordinates. Just as 
the Noether charges do not form supercoordinates neither do the 
respective variations of the Lagrangian. Using the superconformal algebra, 
however, the charges can be assembled into multiplet structures whose components 
transform into each other under a restricted set of SUSY transformations which 
are parametrized by  the anticommuting superspace coordinates $\theta$ , 
$\bar\theta$ themselves rather than by arbitrary anticommuting parameters. Such 
multiplet structures are referred to as quasi-supercoordinates \cite{cps}. The 
restricted supersymmetry relating the Noether charges can also be used to 
connect the various symmetry breaking terms. These 
quasi-supercoordinates of charges (called supercharges) are constructed in 
Appendix B. The quasi-supercoordinates of charges are all constructed from 
the Hamiltonian and SUSY charge supercoordinates \footnote{Note that just as is 
the case in supersymmetric field theory, the Hamiltonian and SUSY charges can be 
reconstructed from the $R$ symmetry charge\cite{cl}.} and the supercoordinate 
charges associated with the 
chiral and antichiral global Weyl scaling of the supercoordinates (that is the 
chiral 
and antichiral number operators).  The $U_J(1)$ symmetry Noether charge forms 
its 
own supercoordinate.  Due to the quasi-supersymmetry covariance, 
the variations of the action under the various symmetry transformations are 
also related.  In fact all superconformal variations of the action are obtained 
from 
combinations of the $U_J(1)$ symmetry variation of the Lagrangian and the 
conformal SUSY variations of the Lagrangian.

Since the Noether construction is a SUSY covariant proceedure, as long as the 
variation of the supercoordinates under consideration is SUSY covariant, the 
resulting Noether charge will be a supercoordinate.  Indeed, the $U_J(1)$ 
transformation of the supercoordinates is a covariant transformation in that it 
commutes with the SUSY variations of the coordinates.  Thus the Noether 
construction of the $U_J(1)$ charge results in a supercoordinate.  Once again, 
it is the $\theta$, $\bar\theta$ independent component which is the actual 
generator, while the time derivative of the higher components simply integrate 
to zero. The supercharge given in equation (\ref{U(1)Q})
\be
J = -2i g D\phi \bar{D}\bar\phi +4g \left(\phi \dot{\bar\phi} -\dot\phi \bar\phi 
\right) -i \left( \phi g_\phi + \bar\phi g_{\bar\phi} \right) D\phi \bar{D} 
\bar\phi +i \left( \phi K_\phi + \bar\phi K_{\bar\phi} \right) 
\label{U(1)Q1}
\ee
can be expanded in components as 
\be
J = J_{00} +\theta \sqrt{2} J_{10} + \bar\theta \sqrt{2} J_{01} 
+\theta\bar\theta J_{11} ,
\label{U(1)J}
\ee
where the $\theta, \bar{\theta}$ independent component is the $U_J(1)$ symmetry 
generator 
\be
J_{00}=j= -4ig\xi\bar\xi -4g(\dot{z}\bar{z} - z \dot{\bar{z}} ) +i(zK_z 
+\bar{z}K_{\bar{z}} )  -2i (zg_z +\bar{z}g_{\bar{z}} )\xi\bar\xi  ,
\ee
while the higher components
\bea
J_{10} &=& - \bar{z}\frac{\delta \Gamma_0}{\delta \bar\xi} -2(zK_z -
\bar{z}K_{\bar{z}})_{z} \xi +4i (zg_z -\bar{z} g_{\bar{z}})\dot{\bar{z}}\xi \cr
J_{01} &=& {z}\frac{\delta \Gamma_0}{\delta \xi} +2(zK_z -
\bar{z}K_{\bar{z}})_{\bar{z}} \bar\xi + 4i (zg_z -\bar{z} 
g_{\bar{z}})\dot{z}\bar\xi \cr
J_{11} &=& -i(z\frac{\delta \Gamma_0}{\delta z} +\bar{z}\frac{\delta 
\Gamma_0}{\delta \bar{z}} + \xi \frac{\delta \Gamma_0}{\delta \xi} + \bar\xi 
\frac{\delta \Gamma_0}{\delta \bar\xi}) \cr
 & & \qquad \qquad +\frac{d}{dt} \left( (zK_z -\bar{z} K_{\bar{z}} ) +2 (zg_z -
\bar{z} g_{\bar{z}} ) \xi\bar\xi \right).
\eea 
do not generate any symmetries.

The $U_J(1)$ charge obeys the conservation equation
\be
\frac{d}{d t} j = \left( - z\frac{\delta \Gamma_0}{\delta z} 
+\bar{z}\frac{\delta \Gamma_0}{\delta \bar{z}} - \xi \frac{\delta 
\Gamma_0}{\delta \xi} + \bar\xi \frac{\delta \Gamma_0}{\delta \bar\xi}\right)  -
i\delta^{U(1)} L ,
\ee
where the $U_J(1)$ variations of the component coordinates are
\bea
\delta^{U(1)} z = iz    ~~ &,&~~  \delta^{U(1)}  \bar{z} = -i\bar{z} \cr
\delta^{U(1)}  \xi = i \xi    ~~&,&~~  \delta^{U(1)}  \bar{\xi} = -i \bar{\xi} .
\eea
The $U_J(1)$ variation of the component coordinate Lagrangian $L$ is 
\bea
\delta^{U(1)} L &=& \left(4i \dot{z}\dot{\bar{z}} +2\xi \bar\xi 
\left(\dot{z}\frac{\partial}{\partial z} -\dot{\bar{z}} \frac{\partial}{\partial 
\bar{z}} \right) -2\left( \xi \dot{\bar\xi} -\dot\xi \bar\xi \right)\right) 
\left(zg_z -\bar{z}g_{\bar{z}}\right) \cr
 & & -\left( \dot{z} \frac{\partial}{\partial z} - 
\dot{\bar{z}}\frac{\partial}{\partial \bar{z}} + i\xi \bar\xi 
\frac{\partial^2}{\partial z \partial \bar{z}} \right) \left( zK_z -\bar{z} 
K_{\bar{z}}\right) 
\eea
which gives the structure of the $U_J(1)$ Ward identity.  In superspace, this 
identity is obtained by applying the action principle to equation (\ref{U(1)Q}) 
yielding
\be
\hat{w}^{U(1)} \Gamma = (w^{U(1)} -\bar{w}^{U(1)} )\Gamma = 
[\frac{\partial}{\partial t} J] \Gamma +\left[\frac{i}{4} [D , \bar{D} ] 
\Delta^{U(1)}{\cal L} \right]\Gamma .
\ee
Integrating over time yields the variation of the effective action
\be
\delta^{U(1)} \Gamma = \int dt \hat{w}^{U(1)} \Gamma = [\frac{i}{2} \int dV 
\Delta^{U(1)}{\cal L}] \Gamma  .
\ee

Similarly, since the time translation of the supercoordinates is a covariant 
transformation, the Noether construction also yields a supercoordinate charge 
whose first component is the Hamiltonian $h$
\bea
H &=& 4g \dot\phi \dot{\bar\phi} + K_{\phi\bar\phi} D\phi \bar{D}\bar\phi \cr
 &=& h +\theta \sqrt{2} H_{10} +\bar\theta \sqrt{2} H_{01} +\theta\bar\theta 
H_{11} .
\eea
Explicitly, the component operators are given by
\bea
h &=& 4g \dot{z} \dot{\bar{z}} +2 K_{z\bar{z}} \xi\bar\xi \cr
H_{10} &=&   \dot{\bar{z}} \frac{\delta \Gamma_0}{\delta \bar\xi}    \cr
H_{01} &=& \dot{z} \frac{\delta \Gamma_0}{\delta \xi}   \cr
H_{11} &=&  \left(-i\dot{z} \frac{\delta \Gamma_0}{\delta z} +i\dot{\bar{z}} 
\frac{\delta \Gamma_0}{\delta \bar{z}} -i\dot\xi \frac{\delta \Gamma_0}{\delta 
\xi} +i\dot{\bar\xi} \frac{\delta \Gamma_0}{\delta \bar\xi} \right)     .
\eea
Time translations are generated by $h$ which satisfies the conservation law
\be
\frac{d}{d t} h = -\left( \dot{z} \frac{\delta \Gamma_0}{\delta z} 
+\dot{\bar{z}} \frac{\delta \Gamma_0}{\delta \bar{z}} +\dot\xi \frac{\delta 
\Gamma_0}{\delta \xi} +\dot{\bar\xi} \frac{\delta \Gamma_0}{\delta \bar\xi} 
\right) .
\ee
Equivalently, applying the quantum action principle to equation (\ref{HQ}) 
yields the local Ward identity
\be
\hat{w}^H \Gamma = (w^H - \bar{w}^H )\Gamma = -i\frac{\partial}{\partial t} H  
\ee
while integrating over time yields the time translation invariance of the 
effective action 
\be
\delta^H \Gamma = \int dt \hat{w}^H \Gamma = 0 .
\ee

The remaining superconformal transformations do not (anti-)commute with the SUSY 
charges.  However, since both the $R$ and SUSY transformations commute with $H$, 
the quasi-supercoordinate construction will lead to a combination of charges 
that also forms a supercoordinate since they act as constant supercoordinates.  
Specifically, the quasi-supercoordinate combination of variations corresponding 
to $\hat{R}$ is nothing but the SUSY covariant $U(1)$ transformation
\bea
i[\hat{R} , \tilde\phi ] &=& i [ R-i\theta Q +i \bar{Q}\bar\theta -
2\theta\bar\theta H, \tilde\phi ] \cr
 &=& \delta^R \tilde\phi -i\theta \delta^Q \tilde\phi -i\bar\theta 
\delta^{\bar{Q}} \tilde\phi -2\theta\bar\theta \delta^H \tilde\phi \cr
 &=& n_\phi \delta^{U(1)} \tilde\phi = \left\{ {in_\phi \phi}\atop{-in_\phi 
\bar\phi} \right.,
\eea
where $\tilde\phi$ is either $\phi$ or $\bar\phi$.  The Noether procedure 
applied to the $U(1)$ symmetry yields a supercoordinate charge as given in 
equation (\ref{U(1)Q}) and (\ref{U(1)Q1}) above.  Expanding the $U(1)$ variation 
of the supercoordinates, multiplying by the corresponding derivative of the 
action, applying the covariant derivatives and inserting it all into one-
particle irreducible functions yields the relation
\bea
-i \left[ w^R \Gamma +i\theta w^Q \Gamma \right.&+& \left. i \bar\theta 
w^{\bar{Q}} \Gamma -2\theta\bar\theta w^H \Gamma \right] = \left[ D \left( 
\delta^R \phi \frac{\delta \Gamma_0}{\delta \phi} \right)\right] \Gamma \cr
 &+& \left[ i\theta D \left( \delta^Q \phi \frac{\delta \Gamma_0}{\delta \phi} 
\right)\right] \Gamma 
 +\left[ i \bar\theta {D} \left( \delta^{\bar{Q}} \phi \frac{\delta 
\Gamma_0}{\delta \phi}\right)\right] \Gamma +\left[-2\theta\bar\theta D \left( 
\delta^H \phi \frac{\delta \Gamma_0}{\delta \phi} \right)\right]\Gamma \cr
 &=&\left[D \left( n_\phi \delta^{U(1)} \phi \frac{\delta \Gamma_0}{\delta \phi} 
\right)\right] \Gamma +  \left[ iD\phi \frac{\delta \Gamma_0}{\delta \phi} 
\right] \Gamma ,
\eea
where the last equality was obtained upon application of the action principle.  
A similar expression is obtained for the antichiral supercoordinate variations 
yielding
\bea
-i\left(\bar{w}^R \Gamma +i\theta \bar{w}^Q \Gamma \right.&+&\left. i \bar\theta 
\bar{w}^{\bar{Q}} \Gamma -2\theta\bar\theta \bar{w}^H \Gamma \right) = 
\left[\bar{D} \left( \delta^R \bar\phi \frac{\delta \Gamma_0}{\delta \bar\phi} 
\right)\right] \Gamma \cr
 &+& \left[ i\theta \bar{D} \left( \delta^Q \bar\phi \frac{\delta 
\Gamma_0}{\delta \bar\phi} \right)\right] \Gamma +\left[i \bar\theta \bar{D} 
\left( \delta^{\bar{Q}} \bar\phi \frac{\delta \Gamma_0}{\delta \bar\phi} 
\right)\right] \Gamma +\left[-2\theta\bar\theta \bar{D} \left( \delta^H \bar\phi 
\frac{\delta \Gamma_0}{\delta \bar\phi} \right)\right] \Gamma \cr
 &=&\bar{D} \left[ n_\phi \delta^{U(1)} \bar\phi \frac{\delta \Gamma_0}{\delta 
\bar\phi} \right] \Gamma - i \left[ \bar{D}\bar\phi \frac{\delta 
\Gamma_0}{\delta \bar\phi} \right] \Gamma  . 
\eea
Taking the difference of these equations and recalling the conservation 
equations for the $U_J(1)$ and $R_0$ charges, equations (\ref{U(1)Q})-
(\ref{RQ2}), the supercoordinate charge $\hat{R}=R= (n_\phi J + R_0)$ is 
obtained and seen to obey the Ward identity equation
\bea
\hat{w}^R \Gamma &=& (w^R -\bar{w}^R ) \Gamma +i\theta (w^Q -\bar{w}^Q )\Gamma 
+i\bar\theta ( w^{\bar{Q}} -\bar{w}^{\bar{Q}} ) \Gamma -2\theta\bar\theta ( w^H 
-\bar{w}^H ) \Gamma \cr
 &=& [- \frac{\partial}{\partial t} \hat{R}]\Gamma +\left[-\frac{i}{2} [ D, 
\bar{D} ] \frac{n_\phi}{2} \Delta^{U(1)} {\cal L}\right] \Gamma  .
\eea
A special case is obtained for $n_\phi = 0$ where the supercoordinate of charges 
$R$ reduces to the conserved $R_0$ charge supercoordinate.  It obeys the Ward 
identity
\be
\hat{w}^{R_0} \Gamma = \frac{\partial}{\partial t} R_0 ,
\ee
where the Ward identity operator is given by $\hat{w}^{R_0} = w^{R_0} - 
\bar{w}^{R_0}$ with

\bea
w^{R_0} &=& iD\phi \frac{\delta}{\delta \phi} \cr
\bar{w}^{R_0} &=& -i\bar{D}\bar\phi \frac{\delta}{\delta \bar\phi} .
\eea

As is discussed in Appendix B, the quasi-supercoordinate $\hat{Q}$ and 
$\hat{\bar{Q}}$ beginning with $Q$ and $\bar{Q}$, respectively, can be 
constructed as
\bea
i [ \hat{Q}, \tilde{\phi} ] &=& i[Q-2i\bar\theta H , \tilde{\phi} ] \cr
 &=&\delta^Q \tilde\phi -2i\bar\theta \delta^H \tilde\phi = D\tilde\phi \cr
i [ \hat{\bar{Q}}, \tilde{\phi} ] &=& i[\bar{Q}+2i\theta H , \tilde{\phi} ] \cr
 &=&\delta^{\bar{Q}} \tilde\phi +2i\theta \delta^H \tilde\phi = 
\bar{D}\tilde\phi .
\label{Qhat}
\eea
Once again there are two equivalent directions along which to proceed.  Since 
$D$ and $\bar{D}$ are covariant derivatives, the Noether construction leads to 
supercoordinate charges $Q$ and $\bar{Q}$.  It is found that
\be
\hat{Q} = {\cal Q}_I^Q -2i\bar\theta {\cal Q}_I^H = -Q  ,
\ee
with the conservation equation
\be
\hat{w}^Q \Gamma = (w^Q - \bar{w}^Q ) \Gamma +2i\bar\theta (w^H - \bar{w}^H ) 
\Gamma = [i\frac{\partial Q}{\partial t} ]\Gamma .
\ee
Likewise
\be
\hat{\bar{Q}} = {\cal Q}_I^{\bar{Q}} +2i\theta {\cal Q}_I^H = -\bar{Q}  ,
\ee
with the corresponding conservation equation
\be
\hat{w}^{\bar{Q}} \Gamma = (w^{\bar{Q}} - \bar{w}^{\bar{Q}}) \Gamma -2i\theta 
(w^H - \bar{w}^H ) \Gamma = [i\frac{\partial \bar{Q}}{\partial t}] \Gamma .
\ee

Equivalently, the Euler-Lagrange equations and hence the Ward identity operators 
can be used directly in order to derive the supercoordinate charges $Q$ and 
$\bar{Q}$.  Taking covariant derivatives of the variations (\ref{Qhat}) gives
\bea
D (\delta^Q \phi \frac{\delta \Gamma_0}{\delta \phi} ) + 2i\bar\theta D ( 
\delta^H \phi \frac{\delta \Gamma_0}{\delta \phi} ) &=& D (D\phi \frac{\delta 
\Gamma_0}{\delta \phi} ) \cr
\bar{D} (\delta^Q \bar\phi \frac{\delta \Gamma_0}{\delta \bar\phi} ) + 
2i\bar\theta \bar{D} ( \delta^H \bar\phi \frac{\delta \Gamma_0}{\delta \bar\phi} 
) &=& -2i\dot{\bar{\phi}} \frac{\delta \Gamma_0}{\delta \bar\phi}  .
\eea
Inserting these identities into the 1-PI functions and using the action 
principle, yields, after their subtraction,
\bea
\hat{w}^Q \Gamma &=& (w^Q - \bar{w}^Q ) \Gamma +2i\bar\theta (w^H - \bar{w}^H ) 
\Gamma \cr
 &=& [D (D\phi \frac{\delta \Gamma_0}{\delta \phi} )]\Gamma + 
[2i\dot{\bar{\phi}} \frac{\delta \Gamma_0}{\delta \bar\phi} ] \Gamma  \cr
 &=& [i\frac{\partial Q}{\partial t} ]\Gamma ,
\eea
where the identities in equations (\ref{SUSYQ1}) and (\ref{SUSYQ2}) have also 
been exploited.  In a  similar fashion, the identities lead to 
\bea
D (\delta^{\bar{Q}} \phi \frac{\delta \Gamma_0}{\delta \phi} ) - 2i\theta D ( 
\delta^H \phi \frac{\delta \Gamma_0}{\delta \phi} ) &=& -2i\dot{\phi} 
\frac{\delta \Gamma_0}{\delta \phi} \cr
\bar{D} (\delta^{\bar{Q}} \bar\phi \frac{\delta \Gamma_0}{\delta \bar\phi} ) - 
2i\theta \bar{D} ( \delta^H \bar\phi \frac{\delta \Gamma_0}{\delta \bar\phi} ) 
&=& [\bar{D} (\bar{D}\bar\phi \frac{\delta \Gamma_0}{\delta \bar\phi} )] .
\eea
Inserting these identities into the 1-PI functions and using the action 
principle produces
\bea
\hat{w}^{\bar{Q}} \Gamma &=& (w^{\bar{Q}} - \bar{w}^{\bar{Q}}) \Gamma -2i\theta 
(w^H - \bar{w}^H ) \Gamma \cr
 &=&[-\bar{D} (\bar{D}\bar\phi \frac{\delta \Gamma_0}{\delta \bar\phi}) ] \Gamma 
+[- 2i\dot{\phi} \frac{\delta \Gamma_0}{\delta \phi} ] \Gamma  \cr
 &=& [i\frac{\partial \bar{Q}}{\partial t} ]\Gamma ,
\eea
where the identities in equations (\ref{SUSYQ1}) and (\ref{SUSYQ2}) have again 
been employed. 

Having considered combinations of transformations that are covariant variations, 
either being the supercoordinate itself or a covariant derivative of the 
supercoordinate, the Noether construction led to a multiplet of charges that was 
a supercoordinate.  Indeed, the multiplet corresponding to the $R_0$ 
transformation contains the supersymmetry charges and the Hamiltonian among its 
components.  Specifically expressing the supercharge $R_0$ in terms of its 
component charges
gives
\bea
R_0 &=& -2ig D\phi \bar{D}\bar\phi \cr
 &=& r_0 +i\theta \sqrt{2} q -i \sqrt{2} \bar{q}\bar\theta +\theta\bar\theta 
\left( 2ih +i (\xi \frac{\delta \Gamma_0}{\delta \xi} +\bar\xi \frac{\delta 
\Gamma_0}{\delta \bar\xi} )\right) ,
\eea
with the component charges given in terms of the component coordinates as
\bea
r_0 &=& -4ig \xi\bar\xi   \cr
q &=& -4g\dot{\bar{z}}\xi  \qquad ;\qquad \bar{q} = 4g\dot{z} \bar\xi \cr
h &=& 4g \dot{z}\dot{\bar{z}} +2K_{z\bar{z}} \xi\bar\xi  .
\eea
The corresponding component charge conservation equations are
\bea
[\frac{\partial}{\partial t} r_0 ]\Gamma &=& [i\left(-\xi \frac{\delta}{\delta 
\xi} + \bar\xi \frac{\delta}{\delta \bar\xi} \right) ]\Gamma \cr
[\frac{\partial}{\partial t} q] \Gamma &=& [i\left(\xi \frac{\delta}{\delta z} -
2i\dot{\bar{z}}\frac{\delta}{\delta \bar\xi} \right)] \Gamma \cr
[\frac{\partial}{\partial t} \bar{q}] \Gamma &=& [i\left(-\bar\xi 
\frac{\delta}{\delta \bar{z}} +2i\dot{z}\frac{\delta}{\delta \xi} \right)] 
\Gamma \cr
[\frac{\partial}{\partial t} h] \Gamma &=& [i\left(\dot{z} \frac{\delta}{\delta 
z} + \dot{\bar{z}} \frac{\delta}{\delta \bar{z}} +\dot\xi \frac{\delta}{\delta 
\xi} + \dot{\bar\xi} \frac{\delta}{\delta \bar\xi} \right) ]\Gamma ,
\eea
where $r_0$ is the $R$ charge for phase transformations involving the Grassmann 
coordinates only, while $q$ and $\bar{q}$ are the SUSY charges and $h$ the 
Hamiltonian.

The remainder of the superconformal charges can only be promoted to quasi-
supercoordinates.  For example, following the procedure outlined in Appendix B, 
the dilatation transformations spawn the quasi-supercoordinate of 
transformations
\bea
i [ \hat{D} , \tilde\phi ] &=& i[ D- \frac{1}{2} \theta Q -\frac{1}{2} 
\bar{Q}\bar\theta , \tilde\phi ] \cr
 &=& \delta^D \tilde\phi -\frac{1}{2} \theta \delta^Q \tilde\phi +\frac{1}{2} 
\bar\theta \delta^{\bar{Q}} \tilde\phi \cr
 &=& d_\phi \tilde\phi + t \partial_t \tilde\phi  ,
\eea
which is not SUSY covariant due to the explicit time dependence.  The Ward 
identity operator corresponding to this transformation is obtained by taking 
covariant derivatives of the above combination of variations.  For the chiral 
supercoordinate, it takes the form
\bea
-iw^{\hat{D}} \Gamma &=& [D ( \delta^D \phi \frac{\delta \Gamma_0}{\delta \phi}) 
]\Gamma + [\frac{1}{2} \theta D ( \delta^Q \phi \frac{\delta \Gamma_0}{\delta 
\phi}) ]\Gamma +[-\frac{1}{2} \bar\theta D ( \delta^{\bar{Q}} \phi \frac{\delta 
\Gamma_0}{\delta \phi}) ]\Gamma \cr
 &=& [\frac{1}{2}  D \phi \frac{\delta \Gamma_0}{\delta \phi} ]\Gamma + [t D 
(\dot{\phi} \frac{\delta \Gamma_0}{\delta \phi}) ]\Gamma + [D( d_\phi \phi 
\frac{\delta \Gamma_0}{\delta \phi}) ]\Gamma 
\eea
while for the transformation of the antichiral supercoordinate 
\bea
-i\bar{w}^{\hat{D}} \Gamma &=& [\bar{D} ( \delta^D \bar\phi \frac{\delta 
\Gamma_0}{\delta \bar\phi} )]\Gamma + [\frac{1}{2} \theta \bar{D} ( \delta^Q 
\bar\phi \frac{\delta \Gamma_0}{\delta \bar\phi} )]\Gamma +[-\frac{1}{2} 
\bar\theta \bar{D} ( \delta^{\bar{Q}} \bar\phi \frac{\delta \Gamma_0}{\delta 
\bar\phi} )]\Gamma \cr
 &=& [\frac{1}{2}  \bar{D} \bar\phi \frac{\delta \Gamma_0}{\delta \bar\phi} 
]\Gamma + [t \bar{D} (\dot{\bar\phi} \frac{\delta \Gamma_0}{\delta \bar\phi}) 
]\Gamma + [\bar{D}( d_\phi \bar\phi \frac{\delta \Gamma_0}{\delta \bar\phi}) 
]\Gamma  . 
\eea
Hence, the Ward identity operator conservation equation becomes
\bea
\hat{w}^D \Gamma &=& (w^{\hat{D}} - \bar{w}^{\hat{D}}) \Gamma \cr
 &=& [-i\frac{\partial}{\partial t} \hat{D}]\Gamma +\left[-\frac{i}{2} [ D, 
\bar{D} ] \Delta^D_{d_\phi} {\cal L}  \right]\Gamma ,
\eea
where the dilatation quasi-supercoordinate of charges $\hat{D}$ is
\be
\hat{D} = tH +d_\phi U .
\ee
Here $U$ is the global Weyl scaling charge given by the supercoordinate
\be
U = 2i ({\cal Q}^N +{\cal Q}^{\bar{N}} ) ,
\ee
while the variation of the Lagrangian is given by
\bea
\Delta^D{d_\phi} {\cal L}  &=& -i\frac{d_\phi}{2} \delta^U {\cal L} + 
\frac{i}{2} R_0 \cr
 &=& \frac{i}{4}(-4id_\phi \delta^N {\cal L}+R_0 ) + \frac{i}{4} (-4i d_\phi 
\delta^{\bar{N}}{\cal L} +R_0 ) .
\eea
This is the same result obtained previously by combining the Noether charges to 
directly form the dilatation quasi-supercoordinate $\hat{D}$ 
\bea
\hat{D} &=& {\cal Q}_I^D +\frac{1}{2} \theta {\cal Q}_I^Q -\frac{1}{2} 
\bar\theta {\cal Q}_I^{\bar{Q}} \cr
 &=& tH + d_\phi U .
\eea

The conformal SUSY Noether charges can be used as the initial components of the 
conformal SUSY quasi-supercoordinates of charges according to 
\bea
\hat{\bar{S}} &=& 2i \left( {\cal Q}_I^{\bar{S}} -\bar\theta \left( 2i {\cal 
Q}_I^D -{\cal Q}_I^R \right)\right) +i\theta\bar\theta {\cal Q}_I^Q  \cr
 &=& tQ +4 g D\phi \bar\phi +2\phi g_\phi D\phi \bar\phi \cr
 &=& tQ +2D(g\phi\bar\phi ) +2i \bar{T}\cr
&=& tQ +4i\bar{T} +2\bar{Y}  \cr
\hat{S} &=& 2i \left( {\cal Q}_I^S +\theta \left( 2i {\cal Q}_I^D +{\cal Q}_I^R 
\right)\right) -i\theta\bar\theta {\cal Q}_I^{\bar{Q}}  \cr
 &=& t\bar{Q} +4 g \phi \bar{D}\bar\phi +2\bar\phi g_{\bar\phi} \phi 
\bar{D}\bar\phi \cr
 &=& t\bar{Q} +2\bar{D}(g\phi\bar\phi ) +2i T \cr
&=&t\bar{Q} +4iT +2Y .
\eea
These correspond to the variations 
\bea
i[\hat{\bar{S}},\tilde\phi] &=& i[\bar{S}+\bar\theta(2iD +R) +i\theta\bar\theta 
Q , \tilde\phi ] \cr
 &=& \delta^{\bar{S}}\tilde\phi +\bar\theta ( 2i \delta^D \tilde\phi +\delta^R 
\tilde\phi ) +i\theta\bar\theta \delta^Q \tilde\phi \cr
 &=& -tD \tilde\phi \cr
i[\hat{S}, \tilde\phi ] &=& i[S +\theta (-2i D + R ) -i\theta\bar\theta \bar{Q} 
, \tilde\phi ] \cr
 &=& \delta^S \tilde\phi +\theta ( -2i \delta^D \tilde\phi +\delta^R \tilde\phi 
) -i\theta\bar\theta \delta^{\bar{Q}} \cr
 &=& -t\bar{D}\tilde\phi .
\eea
Applying the conservation equations for the individual Noether charges gives the 
quasi-supercoordinate conservation equations 
\bea
\hat{w}^{\bar{S}} \Gamma &=& [i\frac{\partial}{\partial t} \hat{\bar{S}}]\Gamma 
+\left[\frac{i}{2} [D, \bar{D} ] \Delta^{\bar{S}} {\cal L}\right] \Gamma +[- 
\bar\theta (\frac{i}{2} [ D, \bar{D} ] ( 2i \Delta^D {\cal L} -\Delta^R {\cal L} 
) )]\Gamma \cr
\hat{w}^S \Gamma &=& [i\frac{\partial}{\partial t} \hat{S}] \Gamma 
+\left[\frac{i}{2} [D, \bar{D} ] \Delta^S  {\cal L}\right] \Gamma + [\theta 
(\frac{i}{2} [ D, \bar{D} ] ( 2i \Delta^D {\cal L} +\Delta^R {\cal L}))]\Gamma  
,
\eea
where the Ward identity operators are 
\bea
\hat{w}^{\bar{S}} &=& w^{\hat{\bar{S}}} -\bar{w}^{\hat{\bar{S}}} \cr
\hat{w}^S &=& w^{\hat{S}} -\bar{w}^{\hat{S}}  ,
\eea
with
\bea
w^{\bar{S}} &=& D( \delta^{\bar{S}} \phi \frac{\delta \Gamma_0}{\delta \phi}) -
\bar\theta D\left((2i \delta^D \phi \frac{\delta \Gamma_0}{\delta \phi} 
+\delta^R \phi \frac{\delta \Gamma_0}{\delta \phi} )\right) +i\theta\bar\theta 
D\left(\delta^Q \phi \frac{\delta \Gamma_0}{\delta \phi}\right) \cr
\bar{w}^{\bar{S}} &=& \bar{D}( \delta^{\bar{S}} \bar\phi \frac{\delta 
\Gamma_0}{\delta \bar\phi}) -\bar\theta \bar{D}\left((2i \delta^D \bar\phi 
\frac{\delta \Gamma_0}{\delta \bar\phi} +\delta^R \bar\phi \frac{\delta 
\Gamma_0}{\delta \bar\phi} )\right) +i\theta\bar\theta \bar{D}\left(\delta^Q 
\bar\phi \frac{\delta \Gamma_0}{\delta \bar\phi}\right) \cr
w^{S} &=& D( \delta^{{S}} \phi \frac{\delta \Gamma_0}{\delta \phi}) -\theta 
D\left((-2i \delta^D \phi \frac{\delta \Gamma_0}{\delta \phi} +\delta^R \phi 
\frac{\delta \Gamma_0}{\delta \phi} )\right) -i\theta\bar\theta 
D\left(\delta^{\bar{Q}} \phi \frac{\delta \Gamma_0}{\delta \phi}\right) \cr
\bar{w}^{S} &=& \bar{D}( \delta^{S} \bar\phi \frac{\delta \Gamma_0}{\delta 
\bar\phi}) -\theta \bar{D}\left((-2i \delta^D \bar\phi \frac{\delta 
\Gamma_0}{\delta \bar\phi} +\delta^R \bar\phi \frac{\delta \Gamma_0}{\delta 
\bar\phi} )\right) -i\theta\bar\theta \bar{D}(\delta^{\bar{Q}} \bar\phi 
\frac{\delta \Gamma_0}{\delta \bar\phi})  .
\eea

Utilizing the identities
\bea
\bar\theta [ D, \bar{D} ] &=& [D, \bar{D} ] \bar\theta +2D \cr
\theta [ D, \bar{D} ] &=& [D, \bar{D} ] \theta +2\bar{D} \cr
\theta\bar\theta [ D, \bar{D} ] &=& [D, \bar{D} ] \theta\bar\theta -[1 - 
2\theta\bar{D} +2\bar\theta \bar{D} ]  ,  
\eea
and employing the  canonical $R$ and scale weights, $n_\phi =-2d_\phi =1 =-
n_{\bar\phi}$, these take the form
\bea
-i\hat{w}^{\bar{S}}\Gamma &=& [\frac{\partial}{\partial t} \hat{\bar{S}}]\Gamma 
+[\frac{1}{2} [D, \bar{D} ] \left(\Delta^{\bar{S}} {\cal L} - \bar\theta  ( 2i 
\Delta^D {\cal L} -\Delta^R {\cal L} )\right)]\Gamma -[D (  2i \Delta^D {\cal L} 
-\Delta^R {\cal L} )] \Gamma \cr
 &=& [\frac{\partial}{\partial t} \hat{\bar{S}}]\Gamma+ [\frac{1}{2} [D, \bar{D} 
] ( 2i \bar{Y} )]\Gamma -[D( 2i\bar\phi K_{\bar\phi} +2i \bar\phi g_{\bar\phi} 
D\phi \bar{D}\bar\phi )]\Gamma\cr
 &=& [\frac{\partial}{\partial t} [ \hat{\bar{S}}-2\bar{Y} ]] \Gamma -
[D(2i\delta^{\bar{N}}{\cal L} +R_0) ]\Gamma \cr
 &=& [\frac{\partial}{\partial t} (tQ)]\Gamma -[Q]\Gamma +2i [\bar\phi 
\frac{\delta \Gamma_0}{\delta \bar\phi} ] \Gamma .
\eea
On the other hand, for general $R$ and scale weights, the Ward identity becomes
\be
-i\hat{w}^{\bar{S}}\Gamma = [\frac{\partial}{\partial t}(tQ) ]\Gamma -[Q]\Gamma 
+[i(n_\phi -2d_\phi ) (\bar\phi \frac{\delta \Gamma_0}{\delta \bar\phi})]\Gamma  
.
\ee
In analogous fashion, for canonical $R$ and scale weights, 
\bea
-i\hat{w}^S \Gamma &=& [\frac{\partial}{\partial t} \hat{S} ]\Gamma +\left[ 
\frac{1}{2} [D, \bar{D} ] \left(\Delta^S {\cal L} + \theta  ( 2i \Delta^D {\cal 
L} +\Delta^R {\cal L} )\right)\right]\Gamma +[\bar{D} (  2i \Delta^D {\cal L} 
+\Delta^R {\cal L} )]\Gamma \cr
 &=& [\frac{\partial}{\partial t} \hat{S}] \Gamma + \left[ \frac{1}{2} [D, 
\bar{D} ] ( -2i Y ) \right]\Gamma+[\bar{D}( 2i\phi K_{\phi} +2i \phi g_{\phi} 
D\phi \bar{D}\bar\phi ) ]\Gamma \cr
 &=& [\frac{\partial}{\partial t} ( \hat{S}-2Y )]\Gamma +[\bar{D} (2i\delta^N 
{\cal L} +R_0 )]\Gamma \cr
 &=& [\frac{\partial}{\partial t} (t\bar{Q})]\Gamma +[-\bar{Q}]\Gamma +[-2i 
(\phi \frac{\delta \Gamma_0}{\delta \phi} ) ]\Gamma   ,
\eea
while for general $R$ and scale weights, the Ward identity takes the form
\be
-i\hat{w}^S\Gamma = [\frac{\partial}{\partial t}(t\bar{Q}) ]\Gamma +[-
\bar{Q}]\Gamma +[-i(n_\phi -2d_\phi ) (\phi \frac{\delta \Gamma_0}{\delta 
\phi})]\Gamma  .
\ee

Finally, the conformal transformation yields the quasi-supercoordinate variation
\bea
i [\hat{K} , \tilde\phi ] &=& i [ K +\theta \bar{S} +S \bar\theta 
+\theta\bar\theta R , \tilde\phi ] \cr
 &=& \delta^K \tilde\phi +\theta \delta^{\bar{S}} \tilde\phi -\bar\theta 
\delta^S \tilde\phi +\theta\bar\theta \delta^R \tilde\phi \cr
 &=& 2t d_\phi \tilde\phi +t^2 \partial_t \tilde\phi  ,
\eea
along with the quasi-supercoordinate of Noether charges
\bea
\hat{K} &=& 2i \left( -{\cal Q}_I^K -\theta {\cal Q}_I^{\bar{S}} +\bar\theta 
{\cal Q}_I^S -\theta\bar\theta {\cal Q}_I^R \right)  \cr
 &=& t( U-tH) -4g\phi\bar\phi  .
\eea
The conservation equation yields the conformal Ward identity
\bea
-i\hat{w}^K \Gamma &=& [\frac{\partial}{\partial t} \hat{K}] \Gamma -
\left[\frac{1}{2} 
 [D, \bar{D} ] \Delta^K {\cal L} \right] \Gamma - \frac{1}{2}\theta \left[[D, 
\bar{D} ]  \Delta^{\bar{S}} {\cal L} \right] \Gamma + \frac{1}{2} \bar\theta 
\left[ [D, 
\bar{D} ]  \Delta^S {\cal L} \right] \Gamma \cr
 & & \qquad\qquad\qquad\qquad - \frac{1}{2} \theta\bar\theta \left[ [D, \bar{D} 
] \Delta^R {\cal L} \right] \Gamma \cr
 &=& \left[\frac{\partial}{\partial t} \hat{K} \right]\Gamma -\left[\frac{1}{2} 
[D, \bar{D} ] \left(\Delta^K {\cal L} +\theta \Delta^{\bar{S}} {\cal L} -
\bar\theta \Delta^S {\cal L} 
+\theta\bar\theta \Delta^R {\cal L} \right)\right] \Gamma \cr
 & & \qquad\qquad\qquad\qquad -\left[\bar{D} \Delta^{\bar{S}} {\cal 
L}\right]\Gamma +\left[D \Delta^S 
{\cal L}\right]\Gamma +\left[ (1-\theta D +\bar\theta \bar{D} ) \Delta^R {\cal 
L} \right]\Gamma ,
\eea
where the conformal Ward identity operator is $\hat{w}^K = w^K -\bar{w}^K$ with 
the chiral and antichiral Ward identity operators given by
\bea
w^K &=& D[\delta^K \phi \frac{\delta \Gamma_0}{\delta \phi} ] -\theta 
D[\delta^{\bar{S}}\phi \frac{\delta \Gamma_0}{\delta \phi} ] +\bar\theta 
D[\delta^S \phi \frac{\delta \Gamma_0}{\delta \phi} ] + \theta\bar\theta D 
[\delta^R \phi \frac{\delta \Gamma_0}{\delta \phi} ] \cr
\bar{w}^K &=& \bar{D}[\delta^K \bar\phi \frac{\delta \Gamma_0}{\delta \bar\phi} 
] -\theta \bar{D}[\delta^{\bar{S}}\bar\phi \frac{\delta \Gamma_0}{\delta 
\bar\phi} ] +\bar\theta \bar{D}[\delta^S \bar\phi \frac{\delta \Gamma_0}{\delta 
\bar\phi} ] + \theta\bar\theta \bar{D} [\delta^R \bar\phi \frac{\delta 
\Gamma_0}{\delta \bar\phi} ] .
\eea
Substituting the expressions for the variations of the Lagrangian from section 3 
and exploiting the identity
\be
[D , \bar{D} ] t = t [D, \bar{D} ] -2i\theta D -2i \bar\theta \bar{D} ,
\ee
the conformal Ward identity is obtained as
\bea
-i\hat{w}^K \Gamma &=& [\frac{\partial}{\partial t} \hat{K} ]\Gamma +[-t 
\frac{i}{4} [D, \bar{D}] (\delta^U {\cal L} +2 R_0  ) ]\Gamma +[-2i(DY +\bar{D} 
\bar{Y} )]\Gamma +[-\frac{1}{2}\Delta^{U(1)} {\cal L} ]\Gamma \cr
 &=& [\frac{\partial}{\partial t} ( \hat{K} + 4 g \phi \bar\phi )]\Gamma +[-
U]\Gamma +[-t\frac{i}{4} [D, \bar{D}] (\delta^U {\cal L} + 2 R_0  ) ]\Gamma \cr
 &=& [\frac{\partial}{\partial t} \left( tU-t^2 H \right)]\Gamma +[-U]\Gamma 
+\left[-t\frac{i}{4} [D, \bar{D} ] ( \delta^U {\cal L} + 2 R_0  ) \right]\Gamma 
.
\eea

In summary, the (quasi-)supercoordinates of charges for the superconformal group 
are given in terms of the Hamiltonian and SUSY charge supercoordinates, which 
are components of the $R$ symmetry charge, and the charges associated with the 
chiral and antichiral global Weyl scaling of the supercoordinates.  In addition, 
the breaking of the superconformal symmetries is given in terms of the breaking 
of these global Weyl scaling symmetries and the additional $U(1)$ $J$ symmetry 
of the Lagrangian.  In particular, the symmetries and conservation equations can 
be summarized as the Ward identities for the effective action:\linebreak

\bigskip
\begin{tabular}{| l | c | l | }
\hline Symmetry & Charge & Ward Identity  \raisebox{-2ex}{\rule{0cm}{5ex}}\\
\hline\hline $U(1)$ & $J$ & $\delta^{U(1)} \Gamma = [\frac{1}{2}\int dV 
\Delta^{U_J(1)} {\cal L} ]\Gamma$   \raisebox{-3ex}{\rule{0cm}{7ex}}\\
\hline Time Translation & $H$ & $\delta^H \Gamma = 0$
\raisebox{-3ex}{\rule{0cm}{7ex}}\\
\hline Supersymmetry & ${Q}$ & ${\hat{\delta}^Q \Gamma = (\delta^Q +2i\bar\theta 
\delta^H)\Gamma =0}$ 
\raisebox{-3ex}{\rule{0cm}{7ex}}\\ \cline{2-3} & ${\bar{Q}}$ & 
${\hat{\delta}^{\bar{Q}} \Gamma = (\delta^{\bar{Q}} -2i\theta \delta^H )\Gamma = 
0}$
\raisebox{-3ex}{\rule{0cm}{7ex}}\\
\hline $R$ Transformation & $R$ & $\hat{\delta}^R \Gamma = (\delta^R +i\theta 
\delta^Q +i\bar\theta \delta^{\bar{Q}} -2\theta\bar\theta \delta^H )\Gamma $
\raisebox{-3ex}{\rule{0cm}{7ex}}\\
 & & $\quad\quad = i[\frac{n_\phi}{2}\int dV \Delta^{U(1)} {\cal L} ]\Gamma $ 
\raisebox{-3ex}{\rule{0cm}{7ex}}\\
\hline Dilatation & $\hat{D}$ & $\hat{\delta}^D \Gamma = (\delta^D +\frac{1}{2} 
\theta \delta^Q -\frac{1}{2}\bar\theta \delta^{\bar{Q}} ) \Gamma = i[\int dV 
\Delta^D_{d_\phi} {\cal L} ]\Gamma $
\raisebox{-3ex}{\rule{0cm}{7ex}}\\
 & & $\quad\quad = -\frac{1}{4}\int dV \left[ (-4id_\phi \delta^{N} {\cal L} 
+R_0 ) + 
(-4id_\phi \delta^{\bar{N}} {\cal L} +R_0 ) \right] \Gamma $
\raisebox{-3ex}{\rule{0cm}{7ex}}\\
\hline Superconformal & $\hat{\bar{S}}$ & $ \hat{\delta}^{\bar{S}} \Gamma = ( 
\delta^{\bar{S}} -\bar\theta [ 2i\delta^D +\delta^R ] +i\theta\bar\theta 
\delta^Q ) \Gamma $
\raisebox{-3ex}{\rule{0cm}{7ex}}\\
 & & $\quad\quad = -i\int dS [-4id_\phi \delta^{\bar{N}} {\cal L} +R_0 ] \Gamma 
$
\raisebox{-3ex}{\rule{0cm}{7ex}}\\ 
\cline{2-3} & $\hat{S}$ & $\hat{\delta}^{S} \Gamma = ( \delta^{S} -\theta [ -
2i\delta^D +\delta^R ] -i\theta\bar\theta \delta^{\bar{Q}} ) \Gamma $
\raisebox{-3ex}{\rule{0cm}{7ex}}\\
 & & $\quad\quad = -i\int d \bar{S} [-4id_\phi \delta^{N} {\cal L} +R_0 ] \Gamma 
$
\raisebox{-3ex}{\rule{0cm}{7ex}}\\
\hline Conformal & $\hat{K}$ & $\hat{\delta}^K \Gamma = ( \delta^K -\theta 
\delta^{\bar{S}} +\bar\theta \delta^S +\theta\bar\theta \delta^R )\Gamma $
\raisebox{-3ex}{\rule{0cm}{7ex}}\\
 & & $\quad = i\int dV [2t\Delta^D_{d_\phi} {\cal L}]\Gamma + i \int dS 
[\Delta^S {\cal L} 
+\theta \frac{n_\phi}{2}\Delta^{U(1)} {\cal L}]\Gamma $\raisebox{-
3ex}{\rule{0cm}{7ex}}\\
 & & $\qquad +i \int d\bar{S} [\Delta^{\bar{S}} {\cal L} +\bar\theta 
\frac{n_\phi}{2}\Delta^{U(1)} {\cal L}]\Gamma 
-i\int dt [\frac{n_\phi}{2} \Delta^{U(1)} {\cal L}] \Gamma $
\raisebox{-3ex}{\rule{0cm}{7ex}}\\
\hline
\end{tabular}
\bigskip

Recall that in the special case that $n_\phi = 0$, the $R$ charge reduces to the 
conserved $R_0$ charge:
\bea
R_0 &=& -2ig D\phi \bar{D}\bar\phi \cr
DR_0 &=& Q \cr
\bar{D} R_0 &=& -\bar{Q} \cr
\frac{1}{2} \left[ D , \bar{D} \right] R_0 &=& 2i H +i \left(D\phi \frac{\delta 
\Gamma_0}{\delta \phi}  - \bar{D}\bar\phi \frac{\delta \Gamma_0}{\delta 
\bar\phi} \right)  \cr
2i\frac{\partial}{\partial t} R_0 &=& \{ D , \bar{D} \} R_0 = -2i \left( D\phi 
\frac{\delta \Gamma_0}{\delta \phi} +\bar{D}\bar\phi \frac{\delta 
\Gamma_0}{\delta \bar\phi} \right) .
\eea
The superconformal Ward identities involve the variance of the Lagrangian under 
the chiral and antichiral global Weyl scaling transformations (these are just 
the chiral and antichiral number operators).  The combinations $(2i\delta^N 
{\cal L} + R_0) = 2i(\phi K_\phi + \phi g_\phi D\phi \bar{D}\bar\phi)$ and 
$(2i\delta^{\bar{N}}{\cal L} +R_0) = 2i(\bar\phi K_{\bar\phi} +\bar\phi 
g_{\bar\phi} D\phi \bar{D}\bar\phi)$ characterize the superconformal symmetry 
breaking.  The strong requirement of superconformal symmetry implies that the 
theory is trivial.  On the other hand, the less restrictive dilatation symmetry 
breaking is characterized by the sum of the above breaking terms: 
$\Delta^D_{d_\phi} {\cal L} =
\frac{id_\phi}{2} \left( \delta^U {\cal L} +2R_0 \right)  
= -d_\phi \left( \left( \phi K_\phi + \bar\phi K_{\bar\phi} \right) + \left( 
\phi g_\phi + \bar\phi g_{\bar\phi} \right) D\phi \bar{D} \bar\phi \right) .$    
As in the case of the superconformal symmetries, the conformal Ward identity is 
satisfied only for a free theory.

\newpage

\setcounter{equation}{0}
\renewcommand{\theequation}{\thenewapp.\arabic{equation}}

\section*{Appendix A: \,\,\,  Superspace and supercoordinates}

$N=1$ supersymmetry (SUSY) is generated by a complex Grassmann valued (hence 
nilpotent) supersymmetry charge, $Q$, its complex conjugate, $\bar{Q}$, and the 
Hamiltonian, $H$.  Together, these charges satisfy the supersymmetry algebra
\bea
\{Q, \bar{Q}\} &=& 2H \cr
[Q, H] = &0& = [\bar{Q}, H] \cr
\{Q, Q\} = &0& = \{\bar{Q}, \bar{Q}\} ~.
\label{SUSYAlg}
\eea
Since multiplying $Q$ and $\bar{Q}$ by a real phase, $Q \rightarrow e^{i\rho}Q$ 
and $\bar{Q} \rightarrow e^{-i\rho}\bar{Q}$, does not alter the algebra, one can 
define an additional charge, $R$,  generating this automorphism and satisfying 
\bea
[R, Q] &=& Q \cr
[R, \bar{Q}] &=& -\bar{Q} \cr
[R, H] &=& 0 ~.
\eea

$N=1$ superspace is obtained by introducing a complex Grassmann coordinate, 
$\theta$, and its complex conjugate, $\bar\theta$, in addition to the time, $t$ 
so that the triplet $(t, \theta, \bar\theta)$ 
denotes points in superspace. \footnote{For the construction of superspace and 
supercoordinates in extended SUSY, see \cite{Hull}.} The representation of the 
symmetry transformations as superspace differential operators on 
supercoordinates is most readily obtained from the motion that group 
multiplication induces on the parameter space of the SUSY graded Lie group.  An 
element of this group is obtained by exponentiation of the sum of the products 
of the charges with the corresponding superspace point giving
\be
G(t, \theta, \bar\theta) = e^{i(tH +\theta Q +\bar{Q}\bar\theta)}= e^{i(\theta Q 
+\bar{Q}\bar\theta)} e^{itH} ~.
\ee
Multiplication of group elements is determined from the algebra (\ref{SUSYAlg}) 
by exploiting the Baker-Campbell-Hausdorff formula. As a consequence of the SUSY 
algebra, the Baker-Campbell-Hausdorff formula truncates to 
\be
e^A e^B = e^{A+B+\frac{1}{2}[A,B]} ~.
\ee
It follows that the multiplication law for group elements is given by
\be
G(t^\prime, \theta^\prime, \bar\theta^\prime) G(t, \theta, \bar\theta) = 
G(t+t^\prime +i(\theta^\prime\bar\theta -\theta\bar\theta^\prime), \theta + 
\theta^\prime, \bar\theta +\bar\theta^\prime) ~.
\ee
Hence, group multiplication (from the left) induces the movement in parameter 
space
$(t, \theta, \bar\theta)\rightarrow (t+t^\prime +i(\theta^\prime\bar\theta -
\theta\bar\theta^\prime), \theta +\theta^\prime, \bar\theta 
+\bar\theta^\prime)$.

A supercoordinate, $\phi (t, \theta, \bar\theta)$, whose transformation 
properties under $H$, $Q$ and $\bar{Q}$ are determined by the above shift, is 
defined so that
\bea
\phi(t, \theta, \bar\theta) &=& G(t, \theta, \bar\theta) \phi (0,0,0) G^{-1}(t, 
\theta, \bar\theta) \cr
 &=& e^{i(\theta Q +\bar{Q}\bar\theta)} \phi (t,0,0) e^{-i(\theta Q 
+\bar{Q}\bar\theta)} ,
\eea
with $\phi (t,0,0)=e^{itH} \phi (0,0,0) e^{-itH}$ .
It then follows that
\be
G(t^\prime, \theta^\prime, \bar\theta^\prime) \phi (t, \theta, \bar\theta) G^{-
1}(t^\prime, \theta^\prime, \bar\theta^\prime) = \phi (t+t^\prime 
+i(\theta^\prime\bar\theta -\theta\bar\theta^\prime), \theta +\theta^\prime, 
\bar\theta +\bar\theta^\prime) ~.
\label{SUSYVariation}
\ee
For infinitesimal superspace parameters $(\epsilon, \eta, \bar\eta)$, the group 
elements and shifted supercoordinate can be Taylor expanded about $(t, \theta, 
\bar\theta)$. Employing the notation $\delta^{\cal Q} \phi$ for the superspace 
differential operator acting on $\phi$ corresponding to the operator ${\cal Q}$ 
so that
\be
i[{\cal Q}, \phi ] = \delta^{\cal Q} \phi ~,
\ee
for ${\cal Q} \in \{H, ~ Q, ~ \bar{Q}\}$, and  recalling that differentiation 
with respect to the Grassmann coordinates is given by
\be
\frac{\partial}{\partial \theta} \theta = 1 \qquad \qquad 
\frac{\partial}{\partial \bar\theta} \bar\theta = 1 ,
\ee
with all other derivatives vanishing, the representation of the SUSY 
transformations as superspace differential operators is secured as
\bea
i[H, \phi ] &=& \frac{\partial}{\partial t} \phi \equiv \delta^H \phi \cr
i[Q, \phi ] &=& \left( \frac{\partial}{\partial \theta} + i\bar\theta 
\frac{\partial}{\partial t}\right) \phi \equiv \delta^Q \phi \cr
i[\bar{Q}, \phi] &=& \left(-\frac{\partial}{\partial \bar\theta} - i\theta 
\frac{\partial}{\partial t}\right) \phi \equiv \delta^{\bar{Q}} \phi ~. 
\eea
From the charge algebra, $\{Q, \bar{Q}\} =2H$, and the identity for nested 
commutators,
$[\{Q, \bar{Q}\}, \phi ] =\{\delta^Q ,\delta^{\bar{Q}}\}\phi$, it is seen that 
the superspace differential operator variations $\delta^{\cal Q}$ obey the same 
algebra (up to a factor of $-i$) as the charges:
\bea
\{\delta^Q , \delta^{\bar{Q}}\}  &=& -2i \delta^H \cr
[\delta^Q, \delta^H ] = &0& = [\delta^{\bar{Q}}, \delta^H ] \cr
\{\delta^Q ,\delta^Q \} = &0& = \{\delta^{\bar{Q}}, \delta^{\bar{Q}}\} ~.
\eea

Since SUSY transformations induce a real time translation, 
$i(\theta^\prime\bar\theta -\theta\bar\theta^\prime)$ in equation 
(\ref{SUSYVariation}), $\phi (t, \theta, \bar\theta)$ can be chosen to be real 
and denoted by $\Phi$.  This real supercoordinate, $\Phi^\dagger = \Phi$, has a 
component expansion   
\be
\Phi (t, \theta, \bar\theta ) = x(t) + i\theta \psi (t) -i \bar\psi (t) 
\bar\theta + \theta\bar\theta W(t) ~.
\ee
The time dependent coefficients in this decomposition consist of the even 
element coordinates $x(t)$ and $W(t)$  and the Grassmann odd elements $\psi (t)$ 
and $\bar\psi (t)$.  The reality of $\Phi$ dictates that $x^\dagger = x$, 
$W^\dagger = W $ and $\bar\psi = \psi^\dagger $. 

The $R$ symmetry variations are determined in an analogous manner using the 
group multiplication law
\be
e^{i\alpha R} G(t, \theta, \bar\theta) e^{-i\alpha R} = G(t, e^{i\alpha}\theta, 
e^{-i\alpha}\bar\theta) ~,
\ee
with $\alpha$ a real parameter. Since $R$ transformations form an Abelian 
subgroup, the variation of $\phi$ is given by
\be
e^{i\alpha R} \phi (t, \theta, \bar\theta) e^{-i\alpha R} = e^{i\alpha n_R}\phi 
(t, e^{i\alpha}\theta, e^{-i\alpha}\bar\theta) ~,
\ee
where $n_R$ is the intrinsic $R$ charge (or $R$-weight) of the $\phi$ 
supercoordinate.
For real $\Phi$, this $R$-weight is required to vanish; $n_R = 0$. On the other 
hand, for the case of complex $\phi$, which we shall consider next, $n_R$ is 
arbitrary.  For infinitesimal $\alpha$, the $R$ transformations yield the $R$ 
variation of the supercoordinate, 
\be
i[R, \phi (t, \theta, \bar\theta)] = i\left( n_R +\theta 
\frac{\partial}{\partial \theta} -\bar\theta 
\frac{\partial}{\partial\bar\theta}\right) \phi (t, \theta, \bar\theta) \equiv 
\delta^R \phi (t, \theta, \bar\theta) ~.
\ee
Once again the charge algebra $[R, Q]=Q$ and $[R,\bar{Q}] =-\bar{Q}$ implies 
\be
[\delta^R, \delta^Q ] = -i \delta^Q \qquad , \qquad [\delta^R, \delta^{\bar{Q}} 
] = i \delta^{\bar{Q}} ~,
\ee
as is verified by direct calculation using the explicit forms for $\delta^R$, 
$\delta^Q$, and $\delta^{\bar{Q}}$.

An alternative induced motion in parameter space can be obtained by group 
multiplication on the right 
\be
G(t, \theta, \bar\theta)G(t^\prime, \theta^\prime, \bar\theta^\prime) = 
G(t+t^\prime -i(\theta^\prime\bar\theta - \theta\bar\theta^\prime ), \theta + 
\theta^\prime, \bar\theta +\bar\theta^\prime ) ~
\ee
giving
\be
(t,\theta,\bar\theta)\rightarrow (t+t^\prime-i(\theta^\prime\bar\theta - 
\theta\bar\theta^\prime),\theta + \theta^\prime, \bar\theta +\bar\theta^\prime ) 
\ee 
For the supercoordinate $\varphi(t, \theta, \bar\theta)$ defined by
\be
\varphi (t, \theta, \bar\theta) = G^{-1}(t, \theta, \bar\theta) \phi (0,0,0) 
G(t, \theta, \bar\theta) = \phi (-t,-\theta, -\bar\theta) ~,
\ee
the induced motion yields
\be
G^{-1}(t^\prime, \theta^\prime, \bar\theta^\prime ) \varphi(t, \theta, 
\bar\theta)G(t^\prime, \theta^\prime, \bar\theta^\prime) = \varphi (t+t^\prime -
i(\theta^\prime\bar\theta - \theta\bar\theta^\prime ), \theta + \theta^\prime, 
\bar\theta +\bar\theta^\prime ) ~.
\ee
For infinitesimal superspace parameters, this right multiplication induced 
motion leads to \bea
-i[H, \varphi ] &=& \frac{\partial}{\partial t} \varphi  \cr
-i[Q, \varphi ] &=& \left( \frac{\partial}{\partial \theta} - i\bar\theta 
\frac{\partial}{\partial t}\right) \varphi \equiv D \varphi \cr
-i[\bar{Q}, \varphi] &=& \left(-\frac{\partial}{\partial \bar\theta} + i\theta 
\frac{\partial}{\partial t}\right) \varphi \equiv \bar{D} \varphi ~. 
\eea
where the covariant Grassmann derivatives are defined as 
\bea
D &=& \frac{\partial}{\partial \theta} - i\bar\theta \frac{\partial}{\partial t} 
\cr
\bar{D} &=& -\frac{\partial}{\partial \bar\theta} + i\theta 
\frac{\partial}{\partial t} ~.
\eea
Note that 
\bea
\{\delta^Q , D \} = &0& = \{\delta^{\bar{Q}}, D\} \cr
\{\delta^Q , \bar{D} \} = &0& = \{\delta^{\bar{Q}}, \bar{D}\} ~,
\eea
while
\bea
\{D, \bar{D}\} &=& 2i\frac{\partial}{\partial t} \cr
\{D , D \} = &0& = \{\bar{D}, \bar{D}\} ~,
\eea
(Since $[\partial /\partial t , \delta^Q ]=0=[\partial /\partial t , 
\delta^{\bar{Q}} ]$, it also follows that $\partial / \partial t$ is a covariant 
derivative).

The covariant derivatives can be used to obtain irreducible representations of 
the SUSY algebra by covariantly constraining the general supercoordinate $\phi$ 
so that either $\bar{D} \phi =0$, producing a chiral supercoordinate, or $D\phi 
=0$, which yields an antichiral supercoordinate. (As discussed below, $\bar{D} 
\phi=0$ implies that the chiral supercoordinate  is necessarily complex and 
distinct from the complex antichiral supercoordinate; i.e. $\phi \neq 
\bar\phi$).  The solution to the chiral constraint, $\bar{D} \phi =0$, is given 
by
\be
\phi (t, \theta, \bar\theta) = e^{-i\theta\bar\theta \frac{\partial}{\partial 
t}}\phi (t, \theta, 0) ~,
\ee
while the antichiral constraint, $D \bar\phi =0$, is satisfied provided
\be
\bar\phi (t, \theta, \bar\theta) = e^{i\theta\bar\theta\frac{\partial}{\partial 
t}} \bar\phi (t, 0, \bar\theta ) ~.
\ee
The component expansions of the chiral and antichiral supercoordinates contain 
complex coordinates $z,~ \bar{z}$ and complex (spinning) Grassmann coordinates 
$\xi,~\bar\xi$ so that
\bea
\phi (t, \theta , \bar\theta) &=& = e^{-i\theta\bar\theta 
\frac{\partial}{\partial t}}\left(z(t) + i \sqrt{2} \theta \xi (t) \right) \cr
\bar\phi (t, \theta , \bar\theta) &=& 
e^{i\theta\bar\theta\frac{\partial}{\partial t}} \left( \bar{z}(t) -i \sqrt{2} 
\bar\xi (t) \bar\theta \right) ~.
\label{chiralcomponents}
\eea

Alternative parametrizations of the group elements are provided by  
\bea
G_1 (t, \theta, \bar\theta) &=& e^{i(tH +\theta Q)} 
e^{i\bar{Q}\bar\theta}=e^{i\theta Q}e^{i\bar{Q}\bar\theta} e^{itH} \cr
G_2 (t, \theta, \bar\theta) &=& e^{i(tH +\bar{Q}\bar\theta )} e^{i \theta Q}= 
e^{i\bar{Q}\bar\theta} e^{i\theta Q} e^{itH}   ~.
\eea
The relation between each parametrization is obtained using 
the Baker-Campbell-Hausdorff formula as
\be
G(t, \theta, \bar\theta) = G_1 (t-i\theta\bar\theta, \theta, \bar\theta ) = G_2 
(t+i\theta\bar\theta, \theta, \bar\theta ) ~.
\ee
The parametrizations differ by purely imaginary shifts, $\pm i 
\theta\bar\theta$, in the time coordinate. Proceeding just as before, left group 
multiplication with these representations also induces a motion in parameter 
space which now take the forms
\bea
G_1 (t^\prime, \theta^\prime, \bar\theta^\prime ) G_1 (t, \theta, \bar\theta) 
&=& G_1 ( t+t^\prime -2i\theta\bar\theta^\prime , \theta +\theta^\prime, 
\bar\theta + \bar\theta^\prime ) \cr
G_2 (t^\prime, \theta^\prime, \bar\theta^\prime ) G_2 (t, \theta, \bar\theta ) 
&=& G_2 (t +t^\prime +2i\theta^\prime\bar\theta, \theta + \theta^\prime, 
\bar\theta +\bar\theta^\prime ) ~.
\eea
It follows that two different supercoordinate representations can be defined 
using $G_1$ and $G_2$ as 
\be
\phi_{1,2}(t, \theta, \bar\theta) = G_{1,2}(t, \theta, \bar\theta) \phi (0,0,0) 
G^{-1}_{1,2}(t, \theta, \bar\theta) ~.
\ee
Since the group elements are related by imaginary shifts in the time, so are the 
supercoordinate representations
\be
\phi (t, \theta, \bar\theta) = \phi_1 (t -i\theta\bar\theta , \theta, 
\bar\theta) = \phi_2 (t +i\theta\bar\theta , \theta, \bar\theta) ~.
\ee
$\phi$ is called the real representation while $\phi_1$ is known as the chiral 
representation and $\phi_2$ as the antichiral representation.  The SUSY group 
transformations are given by the induced parameter transformations
\bea
G_1(t^\prime, \theta^\prime, \bar\theta^\prime ) \phi_1 (t, \theta, \bar\theta) 
G^{-1}_1 (t^\prime, \theta^\prime, \bar\theta^\prime ) &=& \phi_1 (t + t^\prime 
-2i\theta\bar\theta^\prime , \theta + \theta^\prime , \bar\theta + 
\bar\theta^\prime ) \cr
G_2(t^\prime, \theta^\prime, \bar\theta^\prime ) \phi_2 (t, \theta, \bar\theta) 
G^{-1}_2 (t^\prime, \theta^\prime, \bar\theta^\prime ) &=& \phi_2 (t + t^\prime 
+2i\theta^\prime\bar\theta , \theta + \theta^\prime , \bar\theta + 
\bar\theta^\prime ) ~.
\eea
Once again, for infinitesimal superspace parameters $(\epsilon, \eta, \bar\eta 
)$, the SUSY variations in the chiral and antichiral representations can be 
determined.  Using the notation $i[ {\cal Q}, \phi_{1,2}]\equiv (\delta^{\cal 
Q}\phi)_{1,2}$, the variations are given by
\bea
(\delta^H \phi)_1 &=& \frac{\partial}{\partial t} \phi_1 \qquad , \qquad 
(\delta^H \phi)_2 = \frac{\partial}{\partial t} \phi_2 \cr
(\delta^Q \phi)_1 &=& \frac{\partial}{\partial \theta} \phi_1 \qquad , \qquad 
(\delta^Q \phi)_2 = \left(\frac{\partial}{\partial \theta} +2i\bar\theta 
\frac{\partial}{\partial t} \right) \phi_2 \cr
(\delta^{\bar{Q}} \phi)_1 &=&  \left( -\frac{\partial}{\partial 
\bar\theta}+2i\theta \frac{\partial}{\partial t}\right) \phi_1 \qquad , \qquad 
(\delta^{\bar{Q}} \phi)_2 = -\frac{\partial}{\partial \bar\theta} \phi_2  ~.
\eea
In all representations the nilpotent SUSY variations obey $\{\delta^Q 
,\delta^{\bar{Q}}\}= -2i \frac{\partial}{\partial t}$.

Just as was the case using the real representation, the right cosets can be used 
to define the covariant derivatives in the chiral and antichiral 
representations. From the parameter space motion deduced from 
\bea
G_1 (t, \theta, \bar\theta) G_1(t^\prime, \theta^\prime, \bar\theta^\prime ) &=& 
G_1 (t +t^\prime -2i\theta^\prime \bar\theta, \theta +\theta^\prime , \bar\theta 
+ \bar\theta^\prime ) \cr
G_2 (t, \theta, \bar\theta) G_2(t^\prime, \theta^\prime, \bar\theta^\prime ) &=& 
G_2 (t +t^\prime +2i\theta \bar\theta^\prime, \theta +\theta^\prime , \bar\theta 
+ \bar\theta^\prime ) ~,
\eea
the supercoordinate defined as
\be
\varphi_{1,2} (t, \theta, \bar\theta) = G^{-1}_{1,2} (t, \theta, \bar\theta) 
\phi (0,0,0) G_{1,2}(t, \theta, \bar\theta) ~,
\ee
is found to satisfy
\bea
G_1^{-1} (t^\prime, \theta^\prime, \bar\theta^\prime  ) \varphi_{1} (t, \theta, 
\bar\theta) G_1 (t^\prime, \theta^\prime, \bar\theta^\prime ) &=& \varphi_{1} (t 
+ t^\prime -2i\theta^\prime\bar\theta, \theta + \theta^\prime, \bar\theta 
+\bar\theta^\prime ) \cr
G_2^{-1} (t^\prime, \theta^\prime, \bar\theta^\prime ) \varphi_{2} (t, \theta, 
\bar\theta) G_2 (t^\prime, \theta^\prime, \bar\theta^\prime  ) &=& \varphi_{2} 
(t + t^\prime +2i\theta\bar\theta^\prime, \theta + \theta^\prime, \bar\theta 
+\bar\theta^\prime ) ~.
\eea
Finally, for infinitesimal superspace parameters $(\epsilon, \eta, \bar\eta )$, 
these transformations lead to the expressions for the covariant derivatives in 
the chiral and antichiral representations given by
\bea
D_1 \phi_1 &=&  \left( \frac{\partial}{\partial \theta} -2i\bar\theta 
\frac{\partial}{\partial t}\right) \phi_1 \qquad , \qquad D_2 \phi_2 = 
\frac{\partial}{\partial \theta} \phi_2 \cr
\bar{D}_1 \phi_1 &=& -\frac{\partial}{\partial \bar\theta} \phi_1 \qquad , 
\qquad \bar{D}_2 \phi_2
= \left(-\frac{\partial}{\partial \bar\theta} + 2i\theta\frac{\partial}{\partial 
t} \right) \phi_2 ~.
\eea
Note that $\frac{\partial}{\partial t}$ and  $\{D, 
\bar{D}\}=2i\frac{\partial}{\partial t}$ are SUSY covariant derivatives in all 
representations.

The utility of the (anti-)chiral representations follows from the observation 
that when employed, the chiral and antichiral supercoordinate constraints reduce 
to $\bar\theta$ or $\theta$ independence. Thus in the chiral representation
\be
\bar{D}_1 \phi_1 = 0 = -\frac{\partial}{\partial \bar\theta} \phi_1 ~,
\ee
which implies that $\phi_1$ is independent of $\bar\theta$ and can thus be 
expanded as
\be
\phi_1 (t, \theta, \bar\theta ) = z(t) +i \sqrt{2} \theta \xi (t) .
\ee
On the other hand, in the real representation, the chiral supercoordinate is 
obtained from its representaion in the chiral representaion by a simple 
imaginary shift in time so that  
\bea
\phi (t, \theta, \bar\theta ) &=& \phi_1 (t - i\theta\bar\theta , \theta, 
\bar\theta ) \cr
 &=&  e^{-i\theta\bar\theta \frac{\partial}{\partial t}} \left( z(t) 
+i\sqrt{2}\theta\xi(t) \right) ,
\eea
reproducing the result presented in equation (\ref{chiralcomponents}).  
Similarly, the antichiral constraint in the antichiral representation takes the 
form
\be
D_2 \bar\phi_2 = 0 = \frac{\partial}{\partial \theta} \bar\phi_2 ~,
\ee
which implies $\bar\phi_2$ is independent of $\theta$ and thus has the component 
decomposition
\be
\bar\phi_2 (t, \theta, \bar\theta ) = \bar{z}(t) -i\sqrt{2}\bar\xi (t) 
\bar\theta .
\ee
Once again to obtain the real representation for the antichiral supercoordinate 
one need only perform  an imaginary time shift of the opposite sign to that used 
for the chiral
supercoordinate. So doing, one finds
\bea
\bar\phi (t, \theta, \bar\theta ) &=& \bar\phi_2 (t +i\theta\bar\theta , \theta, 
\bar\theta ) \cr
 &=& e^{i\theta\bar\theta \frac{\partial}{\partial t}} \left( \bar{z} (t) -
i\sqrt{2} \bar\xi (t) \bar\theta \right) ,
\eea
just as in equation (\ref{chiralcomponents}).  In an analogous fashion, the 
variations and the covariant derivatives are also related directly through 
imaginary time shifts as
\bea
\delta^{Q, ~\bar{Q}} &=& e^{-i\theta\bar\theta \frac{\partial}{\partial t}} 
\left( \delta^{Q, ~\bar{Q}}\right)_1 e^{i\theta\bar\theta 
\frac{\partial}{\partial t}} \cr
 &=& e^{i\theta\bar\theta \frac{\partial}{\partial t}} \left( \delta^{Q, 
~\bar{Q}}\right)_2 e^{-i\theta\bar\theta \frac{\partial}{\partial t}} \cr
\stackrel{(-)}{D} &=& e^{-i\theta\bar\theta \frac{\partial}{\partial t}} 
\stackrel{(-)}{D_1} e^{i\theta\bar\theta \frac{\partial}{\partial t}} \cr
 &=& e^{i\theta\bar\theta \frac{\partial}{\partial t}} \stackrel{(-)}{D_2} e^{-
i\theta\bar\theta \frac{\partial}{\partial t}} .
\eea

The SUSY transformations of the component coordinates of each supercoordinate 
are most easily determined in each supercoordinate's respective representation.  
For a chiral supercoordinate in the chiral representation, the SUSY variations 
are
\bea
\left(\delta^Q \phi\right)_1 &=& \delta^Q z -i\sqrt{2} \theta \delta^Q \xi \cr
 &=& \frac{\partial}{\partial \theta} \phi_1 = i\sqrt{2} \xi \cr
\left( \delta^{\bar{Q}} \phi \right)_1 &=& \delta^{\bar{Q}} z -i\sqrt{2}\theta 
\delta^{\bar{Q}} \xi \cr
 &=& \left( -\frac{\partial}{\partial \bar\theta} -2i\theta 
\frac{\partial}{\partial t}\right) \phi_1 = -2i\theta \dot{z} .
\eea
Hence the chiral supercoordinate components' SUSY transformations are
\bea
\delta^Q z &=& i\sqrt{2}\xi \qquad , \qquad \delta^Q \xi =0 \cr
\delta^{\bar{Q}} z &=& 0 \qquad , \qquad \delta^{\bar{Q}} \xi = \sqrt{2} \dot{z} 
.
\eea
For an antichiral supercoordinate in the antichiral representation, the SUSY 
variations are
\bea
\left(\delta^Q \bar\phi\right)_2 &=& \delta^Q \bar{z} -i\sqrt{2} \delta^Q 
\bar\xi \bar\theta \cr
 &=&\left( \frac{\partial}{\partial \theta}+2i\bar\theta 
\frac{\partial}{\partial t}\right) \bar\phi_2 = 2i\bar\theta \dot{\bar{z}} \cr
\left( \delta^{\bar{Q}} \bar\phi \right)_2 &=& \delta^{\bar{Q}} \bar{z} -
i\sqrt{2} \delta^{\bar{Q}} \bar\xi \bar\theta \cr
 &=& -\frac{\partial}{\partial \bar\theta} \bar\phi_2 = -i\sqrt{2}\bar\xi .
\eea
Hence the antichiral supercoordinate components' SUSY transformations are
\bea
\delta^Q \bar{z} &=& 0 \qquad , \qquad \delta^Q \bar\xi = -\sqrt{2} 
\dot{\bar{z}} \cr
\delta^{\bar{Q}} \bar{z} &=& -i\sqrt{2}\bar\xi  \qquad , \qquad \delta^{\bar{Q}} 
\bar\xi = 0 .
\eea
Finally for the real supercoordinate in the real representation, the SUSY 
variations are
\bea
\delta^Q \Phi &=& \delta^Q x -i\theta \delta^Q \psi -i\delta^Q \bar\psi 
\bar\theta +\theta\bar\theta \delta^Q W \cr
 &=& \left( \frac{\partial}{\partial \theta} +i\bar\theta 
\frac{\partial}{\partial t} \right) \Phi = i\psi +\bar\theta 
\left(W+i\dot(x)\right) +\theta\bar\theta \dot{\psi} \cr
\delta^{\bar{Q}} \phi &=& \delta^{\bar{Q}} x -i\theta \delta^{\bar{Q}} \psi -
i\delta^{\bar{Q}} \bar\psi \bar\theta +\theta\bar\theta \delta^{\bar{Q}} W \cr
 &=& \left( -\frac{\partial}{\partial \bar\theta} -i\theta 
\frac{\partial}{\partial t} \right) \phi = -i\bar\psi +\theta \left(W-
i\dot(x)\right) +\theta\bar\theta \dot{\bar\psi} .
\eea
Hence the real supercoordinate components' SUSY transformations are
\bea
\delta^Q x &=& i\psi \qquad , \qquad \delta^Q W = \frac{\partial}{\partial t} 
\psi \cr
\delta^Q \psi &=& 0 \qquad , \qquad \delta^Q \bar\psi = i\left( W +i 
\dot{x}\right) \cr
\delta^{\bar{Q}} x &=& -i\bar\psi \qquad , \qquad \delta^{\bar{Q}} W = 
\frac{\partial}{\partial t} \bar\psi \cr
\delta^{\bar{Q}} \psi &=& i\left( W -i \dot{x}\right)  \qquad , \qquad  
\delta^{\bar{Q}} \bar\psi = 0 .
\eea

The last ($\theta\bar\theta$) component of a real supercoordinate transforms as 
a total time derivative, $\delta^Q W =\frac{\partial}{\partial t} \psi$ and 
$\delta^{\bar{Q}} W = \frac{\partial}{\partial t} \bar\psi$. Hence, assuming 
total differentials in time integrate to zero, it follows that $\int dt W$ is 
SUSY invariant; $\delta^Q \int dt W = 0 = \delta^{\bar{Q}} \int dt W$. In the 
superspace framework, to extract the last component of the real supercoordinate, 
one integrates over a vector measure defined as 
\be
\int dV \Phi = \int dt d\theta d\bar\theta \Phi = -\int dt W .
\ee
In obtaining this result, one employs the convention that integration over the 
Grassmann parameters is defined just as differentiation so that 
\bea
d\theta &=& \frac{\partial}{\partial \theta} \cr
d\bar\theta &=& \frac{\partial}{\partial \bar\theta} .
\eea
Since $\delta^Q \int dt W = \delta^{\bar{Q}} \int dt W= 0$, it follows that 
$\delta^Q \int dV \Phi = \delta^{\bar{Q}} \int dV \Phi =0$ and hence is SUSY 
invariant. 

For chiral, $\chi$, and antichiral, $\bar\chi$, Grassmann supercoordinates, the 
last component of each also transforms as a total derivative or is itself 
invariant.  Thus additional SUSY invariant terms can be secured as integrals of  
a (anti-)chiral Grassmann supercoordinate taken over a (anti-)chiral measure 
defined as $\int dS \chi =\int dt d\theta \chi$ and $\int d\bar{S} \bar\chi = 
\int dt d\bar\theta \bar\chi$.
\newpage

\setcounter{newapp}{2}
\setcounter{equation}{0}

\section*{Appendix B:\,\,\,\,  Superconformal Transformations}

The $N=1$ superconformal group algebra can be obtained using the product rules 
of the orthosymplectic group $OSp(2,1)$ whose algebra consists of 8 (charges) 
generators \cite{dhokervinet}.  Three charges generate the familiar conformal 
subgroup $SO(2,1)$ of quantum mechanics.  They consist of the Hamiltonian $H$ 
which generates time translations, the time dilatation charge $D$ and the 
conformal transformation charge $K$.  The remaining 5 charges consist of the 
generator, $R$, of superspace Grassmann (spinning) coordinate rotations, the 
anticommuting SUSY charges, $Q$ and $\bar{Q}$, and the anticommuting special 
(conformal) SUSY charges, $S$ and $\bar{S}$.  An element of $OSp(2,1)$ can be 
represented by matrices having the generic structure \cite{Duval}
\be
\left(
\begin{tabular}{c | c}
$SO(2,1)$ & $Q$ \\
$H,~D,~K$ & $S$ \\
\hline $\bar{Q}$\quad $\bar{S}$ & $R$ 
\end{tabular}
\right)~.
\ee
By examination of the matrix commutators, the $OSp(2,1)$ algebra can be 
extracted as 

\bea
[H, D] &=& iH \qquad [H, K] = 2iD \qquad [D, K] = iK \cr
\{Q, \bar{Q}\} &=& 2H \qquad\qquad\qquad \{S, \bar{S}\} = 2K \cr
[R, Q] &=& Q \qquad\qquad\qquad [R, S] = -S \cr
[R, \bar{Q}] &=& -\bar{Q} \qquad\qquad\qquad [R, \bar{S}] = \bar{S} \cr
[D, Q] &=& -\frac{i}{2} Q \qquad\qquad\quad [D, \bar{Q}] = -\frac{i}{2} \bar{Q} 
\cr
[D, S] &=& \frac{i}{2} S \qquad\qquad\quad [D, \bar{S}] = \frac{i}{2} \bar{S} 
\cr
[K, Q] &=& i\bar{S} \qquad\qquad\qquad [K, \bar{Q}] = iS \cr
[H, S] &=& -i\bar{Q} \qquad\qquad\qquad  [H, \bar{S}] = -iQ \cr
\{Q, S\} &=& -2D -iR \qquad\quad \{\bar{Q}, \bar{S}\} = -2D +iR 
\eea
with all other (anti-)commutators vanishing.

The representation of these transformations on a general supercoordinate $\phi$ 
can be determined by the techniques of Appendix A.  Employing the notation 
$i[{\cal Q}, \phi ] = \delta^{\cal Q} \phi$ for each charge ${\cal Q}$, the 
variations in the real representation are given by \linebreak
\bigskip
\begin{tabular}{| l | l | }
\hline Symmetry & Transformation  \raisebox{-2ex}{\rule{0cm}{5ex}}\\
\hline\hline Time Translation & $ \delta^H \phi (t, \theta, \bar\theta ) = 
\frac{\partial}{\partial t} \phi (t, \theta, \bar\theta ) $
\raisebox{-3ex}{\rule{0cm}{7ex}}\\
\hline Supersymmetry & $\delta^Q \phi (t, \theta, \bar\theta ) = 
\left(\frac{\partial}{\partial \theta} +i\bar\theta \frac{\partial}{\partial t} 
\right) \phi (t, \theta, \bar\theta ) $
\raisebox{-3ex}{\rule{0cm}{7ex}}\\ 
\cline{2-1} & $\delta^{\bar{Q}} \phi (t, \theta, \bar\theta ) = \left(-
\frac{\partial}{\partial \bar\theta} -i\theta \frac{\partial}{\partial t}\right) 
\phi (t, \theta, \bar\theta ) $
\raisebox{-3ex}{\rule{0cm}{7ex}}\\
\hline $R$ Transformation & $\delta^R \phi (t, \theta, \bar\theta ) = i\left( 
n_\phi  +\theta \frac{\partial}{\partial \theta} -\bar\theta 
\frac{\partial}{\partial \bar\theta} \right) \phi (t, \theta, \bar\theta ) $
\raisebox{-3ex}{\rule{0cm}{7ex}}\\
\hline Dilatation & $ \delta^D \phi (t, \theta, \bar\theta ) = \left( d_\phi + 
t\frac{\partial}{\partial t} +\frac{1}{2}\theta \frac{\partial}{\partial \theta} 
+ \frac{1}{2}\bar\theta\frac{\partial}{\partial \bar\theta}\right) \phi (t, 
\theta, \bar\theta ) $
\raisebox{-3ex}{\rule{0cm}{7ex}}\\
\hline Conformal &  $\delta^K \phi (t, \theta, \bar\theta ) = \left( 2td_\phi 
+in_\phi \theta\bar\theta +t\left(\theta \frac{\partial}{\partial \theta} + 
\bar\theta \frac{\partial}{\partial \bar\theta} \right) +t^2 
\frac{\partial}{\partial t} \right) \phi (t, \theta, \bar\theta ) $
\raisebox{-3ex}{\rule{0cm}{7ex}}\\
\hline Superconformal & $ \delta^S \phi (t, \theta, \bar\theta ) = \left( 
i\theta (2d_\phi -n_\phi) + t\frac{\partial}{\partial \bar\theta} +it\theta 
\frac{\partial}{\partial t} +i \theta\bar\theta \frac{\partial}{\partial 
\bar\theta} \right) \phi (t, \theta, \bar\theta ) $
\raisebox{-3ex}{\rule{0cm}{7ex}}\\
\cline{2-1} & $ \delta^{\bar{S}} \phi (t, \theta, \bar\theta ) = \left( -i 
\bar\theta (2d_\phi +n_\phi) - t\frac{\partial}{\partial \theta} -it \bar\theta 
\frac{\partial}{\partial t} +i \theta\bar\theta \frac{\partial}{\partial \theta} 
\right) \phi (t, \theta, \bar\theta ) $
\raisebox{-3ex}{\rule{0cm}{7ex}}\\
\hline
\end{tabular}

\bigskip
\noindent
where $n_\phi$ is the $R$ charge (weight) of the supercoordinate while $d_\phi$ 
is its scaling weight (engineering dimension).  If $\phi$ is hermitean, then 
$n_\phi =0$, while for chiral $\phi$ with $R$-weight $n_\phi$, then the 
antichiral supercoordinate $\bar\phi$ has $R$-weight $n_{\bar\phi} = -n_\phi$.  
Requiring chiral consistency in the commutation relations of the covariant 
derivatives, $D$ and $\bar{D}$, with the transformations imply that for chiral 
supercoordinates  $n_\phi = -2d_\phi $, while  for antichiral supercoordinates 
that $n_{\bar\phi} = 2d_\phi $.  As can be  explicitly demonstrated, the 
variations obey the superconformal algebra. Thus if $[{\cal Q}_1 , {\cal Q}_2 
]_\pm = if {\cal Q}_3$, then $[\delta^{{\cal Q}_1}, \delta^{{\cal Q}_2} ]_\pm = 
f \delta^{{\cal Q}_3}$.

In terms of the component coordinates, the superconformal and $U(1)$ 
transformations are given by \hfill\hfill
\linebreak
\bigskip
\begin{tabular}{| l | l |l |}
\hline Symmetry & Chiral Components & Antichiral Components \raisebox{-
2ex}{\rule{0cm}{5ex}}\\
 & Transformation & Transformation \raisebox{-2ex}{\rule{0cm}{5ex}}\\
\hline\hline Time Translation & $\delta^H z = \dot{z}$ & $\delta^H 
\bar{z} = \dot{\bar{z}}$
\raisebox{-2ex}{\rule{0cm}{5ex}}\\
\cline{2-3} &$\delta^H \xi = \dot{\xi}$ & $\delta^H \bar{\xi} = 
\dot{\bar{\xi}}$
\raisebox{-2ex}{\rule{0cm}{5ex}}\\
\hline Supersymmetry & $\delta^Q z = i\sqrt{2} \xi$ & $\delta^Q \bar{z} 
= 0 $
\raisebox{-2ex}{\rule{0cm}{5ex}}\\ 
\cline{2-3} & $\delta^Q \xi = 0$ & $\delta^Q \bar{\xi} = -
\sqrt{2}\dot{\bar{z}}$
\raisebox{-2ex}{\rule{0cm}{5ex}}\\
\cline{2-3} & $\delta^{\bar{Q}} z = 0$ & $\delta^{\bar{Q}} \bar{z} = -
\bar\xi$
\raisebox{-2ex}{\rule{0cm}{5ex}}\\
\cline{2-3} & $\delta^{\bar{Q}} \xi = +\sqrt{2}\dot{z}$ & 
$\delta^{\bar{Q}} \bar{\xi} = 0$
\raisebox{-2ex}{\rule{0cm}{5ex}}\\
\hline $R_0$ Transformation & $\delta^{R_0} z = 0$ & $\delta^{R_0} 
\bar{z} = 0 $
\raisebox{-2ex}{\rule{0cm}{5ex}}\\
\cline{2-3} & $\delta^{R_0} \xi = +i \xi$ & $\delta^{R_0} \bar{\xi} = -i 
\bar{\xi} $
\raisebox{-2ex}{\rule{0cm}{5ex}}\\
\hline Dilatation & $ \delta^D z = (d + t\partial_t )z$ & $\delta^D 
\bar{z} = (d + t\partial_t )\bar{z}$
\raisebox{-2ex}{\rule{0cm}{5ex}}\\
\cline{2-3} & $\delta^D \xi = (d_\xi + t\partial_t )\xi$ & $\delta^D 
\bar{\xi} = (d_\xi + t\partial_t )\bar{\xi}$
\raisebox{-2ex}{\rule{0cm}{5ex}}\\
\hline Conformal &$\delta^K z = (2td + t^2 \partial_t )z$ & $\delta^K 
\bar{z} = (2td + t^2 \partial_t )\bar{z}$
\raisebox{-2ex}{\rule{0cm}{5ex}}\\
\cline{2-3} & $\delta^K \xi = (2td_\xi + t^2 \partial_t )\xi$ & 
$\delta^K \bar{\xi} = (2td_\xi + t^2 \partial_t )\bar{\xi} $
\raisebox{-2ex}{\rule{0cm}{5ex}}\\
\hline Superconformal & $ \delta^S z = 0$ & $\delta^S \bar{z} = 
it\sqrt{2} \bar\xi$
\raisebox{-2ex}{\rule{0cm}{5ex}}\\
\cline{2-3} & $\delta^S \xi = -\sqrt{2} [2(d_\xi -\frac{1}{2}) +t 
\partial_t ] z$ & $\delta^S \bar{\xi} = 0 $
\raisebox{-2ex}{\rule{0cm}{5ex}}\\
\cline{2-3} & $\delta^{\bar{S}} z = -it\sqrt{2} \xi$ & $\delta^{\bar{S}} 
\bar{z} = 0 $
\raisebox{-2ex}{\rule{0cm}{5ex}}\\
\cline{2-3} & $\delta^{\bar{S}} \xi = 0$ & $\delta^{\bar{S}} \bar{\xi} = 
+\sqrt{2} [2(d_\xi -\frac{1}{2}) + t \partial_t ] \bar{z}$
\raisebox{-2ex}{\rule{0cm}{5ex}}\\
\hline $U_J(1)$ & $\delta^J z = +iz$ & $\delta^J \bar{z} = -i\bar{z} $
\raisebox{-2ex}{\rule{0cm}{5ex}}\\
\cline{2-3} & $\delta^J \xi = +i \xi$ & $\delta^J \bar{\xi} = -i 
\bar{\xi}$
\raisebox{-2ex}{\rule{0cm}{5ex}}\\
\hline
\end{tabular}

\bigskip
\noindent
Note that in order for the superconformal algebra to close, it is necessary that 
$d_\xi = d_\phi + \frac{1}{2}$ .

Some of the superspace dependent charges of the superconformal algebra have 
explicit time dependence and thus do not transform as supercoordinates under 
SUSY transformations. It is possible, however, to add explicit $\theta$ and 
$\bar\theta$ terms to these charges such that the new charge transforms under a 
restricted SUSY transformation as would a time independent supercoordinate while 
still obeying the same superconformal algebra as the associated charge without 
the explicit $\theta$ and $\bar\theta$. Such a quasi-supercoordinate $\hat{{\cal 
Q}}$ is defined so that under a SUSY transformation parametrized with the 
superspace coordinates $\theta$ and $\bar\theta$, it transforms as would a 
supercoordinate so that \cite{cps}
\be
i[\theta Q +\bar{Q}\bar\theta , \hat{{\cal Q}} ] = \left(\theta 
\frac{\partial}{\partial \theta} +\bar\theta \frac{\partial}{\partial 
\bar\theta} \right) \hat{{\cal Q}} .
\ee
The solution to this equation is given by
\be
\hat{{\cal Q}}(\theta, \bar\theta) = e^{i(\theta Q +\bar{Q}\bar\theta)} {\cal Q} 
e^{-i(\theta Q + \bar{Q}\bar\theta)} .
\ee
For each of the superconformal charges the corresponding quasi-supercoordinate 
is constructed 
\bea
\hat{H} &=& H \cr
\hat{Q} &=& Q -2i\bar\theta H \cr
\hat{\bar Q} &=& \bar{Q} +2i\theta H \cr
\hat{R} &=& R-i\theta Q +i \bar{Q}\bar\theta -2\theta\bar\theta H \cr
\hat{D} &=& D -\frac{1}{2} \theta Q -\frac{1}{2}\bar{Q}\bar\theta \cr
\hat{K} &=& K +\theta \bar{S} +S \bar\theta +\theta\bar\theta R \cr
\hat{S} &=& S +\theta (-2iD +R) -i\theta\bar\theta \bar{Q} \cr
\hat{\bar S} &=& \bar{S} +\bar\theta (+2iD +R) +i\theta\bar\theta Q  .
\label{supercharges}
\eea
The quasi-supercoordinates of charges $\hat{\cal Q}$ obey the same 
superconformal algebra as the ${\cal Q}$.
\newpage

\setcounter{newapp}{3}
\setcounter{equation}{0}

\section*{Appendix C:\,\,\,\,  Quantum Action Principle}

The generating functional, $Z[\eta, \bar\eta]$, for ground state expectation 
values of the time ordered products of chiral and antichiral supercoordinates is 
given by the Feynman path integral 
\bea
Z[\eta, \bar\eta] &=& <0| T e^{i\int dS \eta \phi + i\int d\bar{S} \bar\eta 
\bar\phi} |0> \cr
 &=& \int [d\phi][d \bar\phi ] e^{i\left( \Gamma_0 +\int dS \eta \phi +\int d 
\bar{S} \bar\eta \bar\phi \right)} ,
\eea
where $\Gamma_0 = \int dV {\cal L} = \int dt L$ is the classical action.  For 
the model discussed in the body of this paper, the action is given in equation 
(\ref{classicalaction}) or equivalently equation (\ref{classicalactioncomp}). 
Here $\eta$ and $\bar\eta$ are chiral and antichiral Grassmann scalar 
supercoordinate sources having the component decompositions
\bea
\eta (t, \theta, \bar\theta ) &=& e^{-i\theta\bar\theta \frac{\partial}{\partial 
t}}\left( \frac{i}{\sqrt{2}} \chi (t) +\theta J (t) \right) \cr
\bar\eta (t, \theta, \bar\theta ) &=& e^{i\theta\bar\theta 
\frac{\partial}{\partial t}} \left( \frac{i}{\sqrt{2}}  \bar\chi (t) +\bar\theta 
\bar{J} (t) \right) .
\eea
Using the chiral and antichiral measures, $dS =dt \frac{\partial}{\partial 
\theta}$, and  $d\bar{S} = dt \frac{\partial}{\partial \bar\theta}$ 
respectively, the generating functional expressed in terms of the component 
coordinates is 
\be
Z[\eta, \bar\eta] = Z[J, \bar{J}, \chi, \bar\chi] = \int 
[dz][d\bar{z}][d\xi][d\bar\xi ] e^{i\int dt \left\{L + Jz + \bar{J} \bar{z} + 
\chi \xi + \bar\chi \bar\xi \right\}}  .
\ee

The connected function generating functional, $Z^c [\eta, \bar\eta]$, is defined 
as
\be
Z [\eta, \bar\eta ] = e^{Z^c [\eta, \bar\eta ]} .
\ee
The quantum effective action $\Gamma [\phi, \bar\phi ]$, which is the generating 
functional for one-particle irreducible (1-PI) functions is defined as the 
Legendre transform of $Z^c$:
\be
\Gamma [\phi , \bar\phi ] = Z^c [\eta, \bar\eta ] -i\int dS \eta\phi -i \int 
d\bar{S} \bar\eta \bar\phi ,
\ee
where the sources for the 1-PI functions are  the classical supercoordinates 
$\phi$ and $\bar\phi$, defined as
\be
\frac{1}{i}\frac{\delta Z^c}{\delta \eta} = \phi \qquad , \qquad 
\frac{1}{i}\frac{\delta Z^c}{\delta \bar\eta} = \bar\phi .
\ee
In terms of the component coordinates, the Legendre transform is 
\be
\Gamma [z, \bar{z}, \xi, \bar\xi ] = Z^c [J, \bar{J}, \chi, \bar\chi ] -i \int 
dt \left( Jz + \bar{J} \bar{z} + \chi \xi + \bar\chi \bar\xi  \right) ,
\ee
with the classical supercoordinates given by
\bea
\frac{1}{i}\frac{\delta Z^c}{\delta J} &=& z \qquad , \qquad 
\frac{1}{i}\frac{\delta Z^c}{\delta \bar{J}} = \bar{z} \cr
\frac{1}{i}\frac{\delta Z^c}{\delta \chi} &=& \xi \qquad , \qquad 
\frac{1}{i}\frac{\delta Z^c}{\delta \bar{\chi}} = \bar{\xi} .
\eea
It should be noted that in the supercoordinate case, the chiral supercoordinate 
functional derivatives are Grassmann odd. Thus they have been used in the 
Grassmann odd chiral and antichiral integrals in order to give chiral, 
$\delta^S$, and antichiral, $\delta^{\bar{S}}$, superspace delta functions that 
then integrate to Grassmann even functions so that 
\bea
\frac{\delta \phi (1)}{\delta \phi (2)} &=& \delta^S (1,2) = \bar{D}_1 \delta 
(1,2) \cr
 &=& \theta_{12} e^{-i\theta_2\bar\theta_{12} \frac{\partial}{\partial 
t_1}}\delta (t_1-t_2) \cr
\frac{\delta \bar\phi (1)}{\delta \bar\phi (2)} &=& \delta^{\bar{S}} (1,2) = D_1 
\delta (1,2) \cr
&=& \bar\theta_{12 }e^{i\theta_{12}\bar\theta_{2} \frac{\partial}{\partial 
t_1}}\delta (t_1-t_2)  ,
\eea
where the real measure delta function is defined as $\delta (1,2) = \delta (t_1-
t_2) \theta_{12}\bar\theta_{12}$ with $\theta_{12}=\theta_1 - 
\theta_2~~;~~\bar{\theta}_{12}=\bar{\theta}_1-\bar{\theta}_2$.

The supercoordinate Euler-Lagrange equations for the above chiral 
supercoordinate model with action (\ref{classicalaction}) can be derived either 
graphically or by using a formal change of integration variable in the path 
integral. So doing, one finds that, for this action, no renormalization is 
needed. Consequently, the equations of motion for the Green's function 
generating functional are
\be
\frac{\delta \Gamma_0}{\delta \phi (1)} Z[\eta, \bar\eta]  = -\eta (1) Z[\eta, 
\bar\eta] \qquad , \qquad  \frac{\delta \Gamma_0}{\delta \bar\phi (1)} Z[\eta, 
\bar\eta]  = -\bar\eta (1) Z[\eta, \bar\eta] ,
\ee
while for connected functions one finds
\be
\frac{\delta \Gamma_0}{\delta \phi (1)} Z^c[\eta, \bar\eta]  = -\eta (1) \qquad 
, \qquad  \frac{\delta \Gamma_0}{\delta \bar\phi (1)} Z^c[\eta, \bar\eta]   = -
\bar\eta (1) .
\ee
The functional derivatives of the classical action, $\Gamma_0$, are just the 
supercoordinate generalization of the classical Euler-Lagrange derivatives of 
the Lagrangian.  For the action under consideration, $\Gamma_0 = \int dV [ g 
D\phi \bar{D}\bar\phi + K] \equiv \int dV {\cal L}$,
the Euler-Lagrange derivatives are
\bea
\frac{\delta \Gamma_0}{\delta \phi (1)} &=& -\bar{D} \left( \frac{\partial {\cal 
L}}{\partial \phi} -D\left(\frac{\partial {\cal L}}{\partial 
D\phi}\right)\right) \cr
 &=& \bar{D} \left( gD\bar{D} \bar\phi - K_\phi \right) \cr
\frac{\delta \Gamma_0}{\delta \bar\phi (1)} &=& -{D} \left( \frac{\partial {\cal 
L}}{\partial \bar\phi} -\bar{D}\left(\frac{\partial {\cal L}}{\partial 
\bar{D}\bar\phi}\right)\right) \cr
 &=& -D \left( g\bar{D}{D} \phi + K_{\bar\phi} \right) .
\eea
The operator inserted Euler-Lagrange equations can then be Legendre transformed 
in order to obtain the effective action equations of motion
\be
\left[i\frac{\delta \Gamma_0}{\delta \phi (1)}\right]\Gamma [\phi, \bar\phi ] = 
\frac{\delta \Gamma [\phi, \bar\phi]}{\delta \phi (1)} \qquad , \qquad \left[ 
i\frac{\delta \Gamma_0}{\delta \bar\phi (1)}\right] \Gamma [\phi, \bar\phi ] = 
\frac{\delta \Gamma [\phi, \bar\phi]}{\delta \bar\phi (1)} .
\ee

In  a similar fashion, using either graphical or path integral techniques, 
composite operator, ${\cal O}$, inserted Euler-Lagrange equations of motion can 
be gleaned and take the form
\be
\left[ i{\cal O} \frac{\delta \Gamma_0}{\delta \tilde\phi (1)}\right] \Gamma 
[\phi, \bar\phi ] = \frac{\delta }{\delta \tilde\phi (1)} ({\cal O}\Gamma [\phi, 
\bar\phi])~,
\ee
where $\tilde{\phi}$ is either $\phi $ or $\bar{\phi}$. Thus the quantum action 
principle relates the variation of the effective action to that of the classical 
action
\bea
\delta \Gamma &=& \left( \int dS \delta \phi \frac{\delta}{\delta \phi}  + \int 
d\bar{S} \delta \bar\phi \frac{\delta}{\delta \bar\phi} \right) \Gamma \cr
 &=& [i\delta \Gamma_0 ]\Gamma \cr
&=& \left[ i  \int dS \delta \phi \frac{\delta \Gamma_0}{\delta \phi}  + i\int 
d\bar{S} \delta \bar\phi \frac{\delta \Gamma_0}{\delta \bar\phi} \right] \Gamma 
\cr
 &=& \left[ i\int dV \delta {\cal L} \right] \Gamma ,
\eea
where $\delta \phi$ and $\delta\bar\phi$ are chiral and antichiral products of 
the supercoordinates and their derivatives.

\newpage

\setcounter{newapp}{4}
\setcounter{equation}{0}

\section*{Appendix D:\,\,\,\,  Anticommuting Grassmann (Spinning) Coordinate 
Determinant}

In this appendix, we shall explicitly integrate the anticommuting Grassmann 
(spinning) coordinate out of the path integral representation of the Green 
function generating functional resulting in a path integral involving only the 
particle ordinary spatial coordinates. Expressed in terms of the component 
coordinates, including the anticommuting Grassmann (spinning) coordinates, the 
path integral is given by 
\be
Z[J, \bar{J}, \chi, \bar\chi] = \int [dz][d\bar{z}][d\xi][d\bar\xi ] e^{i\int dt 
\left\{L + Jz + \bar{J} \bar{z} + \chi \xi + \bar\chi \bar\xi \right\}}  ~,
\ee
where the Lagrangian with canonically normalized kinetic energy term so  
$g=\frac{1}{4}$ is
\be
L = \dot{z}\dot{\bar{z}} +\frac{i}{2}\left( \xi\dot{\bar\xi} - \dot\xi 
\bar\xi\right) +i\left( K_z\dot{z} -K_{\bar{z}}\dot{\bar{z}}\right) -
2K_{z\bar{z}} \xi \bar\xi  ~.
\ee

Since this Lagrangian is bilinear in the Grassmann (spinning) coordinates, they 
can be integrated out as\cite{GildPat}\cite{CooperFreedman}
\be
\int [d\xi] [d \bar\xi ] e^{i\int dt [{L}_f + \chi \xi +\bar\chi \bar\xi ]}= 
\frac{{\rm Det}[S_F^{-1}]}{{\rm Det}[S_{F0}^{-1}]} e^{i\int dt dt^\prime 
\bar\chi (t) S_F (t-t^\prime ) \chi  (t^\prime ) },
\ee
where \be
{L}_f = \frac{i}{2}(\xi \dot{\bar\xi} -\dot\xi \bar\xi ) -2K_{z\bar{z}} \xi 
\bar\xi  
\ee
is that part of the Lagrangian having dependence on the anticommuting (spinning) 
coordinates $\xi$ and $\bar\xi$. Here we normalized the determinant by its 
noninteracting  ($K_{z\bar{z}}=0$ value so that the path integral gives unity in 
the absence of all interactions and sources. The inverse fermion propagator, 
determined from ${\cal L}_f$, is 
\be
S_F^{-1}(t-t^\prime; z, \bar{z})= \left( i \frac{d}{dt} -2K_{z\bar{z}} 
+i\epsilon \right) \delta (t - t^\prime ) ,
\ee
while $S_{F0}^{-1} = ( i \frac{d}{dt} +i\epsilon ) \delta (t - t^\prime ) $.  
The determinant can be secured as a product of the eigenvalues, $\lambda$, of 
$S_F^{-1}$ defined through the eigenvalue equation
\be
\int dt^\prime S_F^{-1} (t-t^\prime; z, \bar{z} ) \psi_\lambda (t^\prime; z, 
\bar{z}) = \lambda \psi_\lambda (t; z, \bar{z}) .
\ee
The solution to this integral equation is given by 
\be
\psi_\lambda (t; z, \bar{z}) = \psi_\lambda(t_0; z, \bar{z}) e^{-i\lambda (t-
t_0) -2i \int_{t_0}^t d\tau  K_{z\bar{z}}(z,\bar{z})}.
\ee
Eventually, we will want to take the adiabatic limit, $t\rightarrow \infty$. 
Since the time integral in the exponential diverges in this limit, we must 
regulate the integral in the intermediate steps. We do so by introducing a 
cutoff at time $T$. As such, we need to impose boundary conditions on the 
solution. Imposing anti-periodic boundary conditions, the 
anticommuting coordinate eigenfunctions are then required to satisfy
\be
\psi_\lambda (-T) = - \psi_\lambda (T) ~.
\ee
This, in turn, restricts the allowed eigenvalues to be 
\be
\lambda \rightarrow \lambda_n = \frac{(2n +1) \pi}{2 T} -\frac{1}{T} \int_{-
T}^{T} d\tau {K_{z\bar{z}}} ,
\ee
where $n = 0, \pm 1 , \pm 2 , \ldots $ is any integer.  

The desired ratio of determinants is then simply given by the product of the 
eigenvalues as 
\bea
\frac{{\rm Det}[S_F^{-1}]}{{\rm Det}[S_{F0}^{-1}]} &=& \frac{\prod_{n=-
\infty}^\infty \lambda_n}{\prod_{n=-\infty}^\infty 
{\lambda_n}|_{K_{z\bar{z}}=0}}\cr
&=&\prod_{n=-\infty}^{\infty} \left[ 1 - \frac{ \int_{-T}^{T} d\tau 
2{K_{z\bar{z}}}}{(2n+1)\pi} \right] \cr
 &=& \cos{[\int_{-T}^{T} d\tau {K_{z\bar{z}}} ]}  .
\eea
Taking the adiabatic limit, $T\rightarrow \infty$, Green function generating 
functional (ignoring the Grassmann sources for simplicity) takes the form
\be
Z[J, \bar{J}, 0,0] = \int [dz] [d\bar{z}] \left ( e^{i\int dt K_{z\bar{z}}(z, 
\bar{z})} + e^{-i\int dt K_{z\bar{z}}(z, \bar{z})}\right) e^{i\int dt [ 
\dot{z}\dot{\bar{z}} +i(K_z \dot{z} -K_{\bar{z}}\dot{\bar{z}}) +Jz +\bar{J} 
\bar{z} ]}.
\ee
Finally, letting $z=\sqrt{\frac{m}{2}}(x+iy)$,  $\bar{z}=\sqrt{\frac{m}{2}}(x-
iy)$, and scaling $K \rightarrow qK$, with $q$ the particle electric cahrge, so 
that $K_{z\bar{z}}=-\frac{q}{2m}B(x,y)=-\frac{q}{2m}\epsilon_{ij}\partial_i 
A_j(x,y)$, where $B(x,y)$ is the magnetic field
and $\vec{A}$ is the vector potential, we secure the path integral 
representation
\be
Z[J_x, J_y, 0,0] = \int [dx] [dy] \left ( e^{i\int dt \frac{qB(x,y)}{2m}} + e^{-
i\int dt \frac{qB(x,y)}{2m}}\right) e^{i\int dt 
[\frac{1}{2}m\vec{v}^2+q\vec{A}\cdot\vec{v}+J_x x+J_y y ]}.
\ee
Here $\vec{v}=\hat{e}_1\dot{x}+\hat{e}_2\dot{y}$ is the particle velocity and we 
have replaced the complex sources $J, \bar{J}$ with the real sources $J_x, J_y$. 
This path integral expression is completely equivalent (note in this appendix, 
we have set $\hbar =1$) to the path integral resulting from the Lagrangian 
(\ref{Lff}) or the Hamiltonian (\ref{Ham}) where the trace over the Pauli matrix 
magnetic dipole moment interaction Hamiltonian has been taken to obtain the two 
magnetic field exponential terms in the path integral.

\bigskip

\noindent
This work was supported in part by the U.S. Department 
of Energy under grant DE-FG02-91ER40681 (Task B).

\end{document}